\documentclass[aps,prc,twocolumn,superscriptaddress]{revtex4-1}

\usepackage{amsmath}
\usepackage{amsfonts}
\usepackage{graphicx}
\usepackage[caption=false]{subfig}
\usepackage{amssymb}
\usepackage{color}
\usepackage{hyperref}
\usepackage{url}
\usepackage{tikz}
\usepackage{blkarray}
\usepackage{mathpazo}

\hypersetup{
 colorlinks,
 citecolor=black,
 filecolor=black,
 linkcolor=black,
 urlcolor=black}

\newcommand{\ket}[1]{| #1 \rangle}
\newcommand{\bra}[1]{\langle #1 |}
\newcommand{\braket}[1]{\langle #1 \rangle}
\newcommand{\Ho}{\text{Ho}} 
\newcommand{\Dy}{\text{Dy}}   

\usetikzlibrary{decorations.markings}
\usetikzlibrary{decorations.pathmorphing}

\tikzset{wiggle/.style={decorate, decoration=snake}}
\tikzset{->-/.style={decoration={
markings,
mark=at position 0.6 with {\arrow{latex}}},postaction={decorate}}}
\tikzset{dArrow/.style={decoration={
markings,
mark=at position 0.7 with {\arrow{latex}}},postaction={decorate}}}

\begin{document}

\title{\textit{Ab initio} calculation of the electron capture spectrum of $^{163}$Ho:\\Auger-Meitner decay into continuum states}

\author{M. Bra\ss}
\affiliation{Institute for Theoretical Physics, Heidelberg University, Philosophenweg 19, 69120 Heidelberg, Germany}

\author{M. W. Haverkort}
\affiliation{Institute for Theoretical Physics, Heidelberg University, Philosophenweg 19, 69120 Heidelberg, Germany}

\date{\today}

\begin{abstract}
	Determining the electron neutrino mass by electron capture in $^{163}$Ho relies on an accurate understanding of the differential electron capture nuclear decay rate as a function of the distribution of the total decay energy between the neutrino and electronic excitations. The resulting spectrum is dominated by resonances due to local atomic multiplet states with core holes. Coulomb scattering between electrons couple the discrete atomic states, via Auger-Meitner decay, to final states with free electrons. The atomic multiplets are above the auto-ionisation energy, such that the delta functions representing these discrete levels turn into a superposition of Lorentzian, Mahan- and Fano-like line-shapes. We present an \textit{ab initio} method to calculate nuclear decay modifications due to such processes. It includes states with multiple correlated holes in local atomic orbitals interacting with unbound Auger-Meitner electrons. A strong energy-dependent, asymmetric broadening of the resonances in good agreement with recent experiments is found. We present a detailed analysis of the mechanisms determining the final spectral line-shape and discuss both the Fano interference between different resonances, as well as the energy dependence of the Auger-Meitner Coulomb matrix elements. The latter mechanism is shown to be the dominant channel responsible for the asymmetric line-shape of the resonances in the electron capture spectrum of $^{163}$Ho.

\end{abstract}

\pacs{}

\maketitle

\section{\label{sec:Intro}Introduction}
	Despite observing only neutrinos with spin anti-parallell to the neutrino momentum, extensive studies of neutrino flavour oscillations imply that neutrinos are massive particles. These oscillations provide information on the mass difference between the mass eigenstates. The question for the actual values of the masses remains unanswered, as so far only upper bounds have been found \cite{Katrin2019}. One way to tackle this question is studying nuclear decay spectra \cite{Fermi34}. A particular experimentally accesible case for the determination of the neutrino mass from a nuclear decay spectrum is the electron capture (EC) spectrum of $^{163}$Ho \cite{DeRujula:1982us,Alpert:2015gi,Croce:2016dp,Gastaldo:2017ch}. 
	
	By electron capture $^{163}$Ho decays into an excited $^{163}$Dy state while an electron neutrino is emitted. The excited daughter atom undergoes subsequent decay via multiple channels leading to an energy spectrum rich of interesting structures. The endpoint of this spectrum is determined by the energy difference of Ho and Dy ground-states minus the rest mass of the created neutrino. In order to obtain a spectrum that is sensitive to neutrino masses in the sub eV regime high statistics and high resolution data are needed. At the same time a precise theoretical understanding of the spectral features, i.e. the differential electron capture nuclear decay rate, with as little parameters as possible is essential. 
	
	Previous theoretical studies discussed the importance of spectral shake-up and shake-off features \cite{Faessler:2015dg, Faessler:2015ck, Robertson:2015dg, DeRujula:2016cp, Faessler:2017hq, Gastaldo:2017ch}. These calculations show that an even further improved description of the EC spectrum is necessary to explain the experimental data available at that time \cite{Ramitzsch17}. 

	In Ref. \onlinecite{Brass:2018} the sharp features as well as some of the apparent line broadening observed in the experimental spectra is explained by treating the full Coulomb interaction between the electrons restricted to a basis of bound-orbitals. The full quantum mechanical scattering between electrons firstly generates a multiplet splitting of the resonances. This leads to an apparently broader line-width at the resonance, as several resonances with small energy differences are not resolved with the current experimental resolution. Secondly, Coulomb scattering results in additional satellite structures emerging due to bound states with two correlated holes in inner atomic orbitals. 
	
	In a recent experimental work \cite{Clemens:2019} with higher statistics it was shown that the atomic multiplet resonances present in the EC spectrum of Ho have an asymmetric line-shape. This yields noticeably more intensity outside the region of the resonances as one would expect from locally bound states broadened by a Lorentzian spectral-function. 
	
	In the present paper we address the spectral line-broadening due to the emission of electrons subsequent to an electron capture event. Just after such an event the Dy atom is in an excited state composed of one core hole created in the many-electron ground-state wave-function of a Ho atom. This state is not an eigen-state of the system and thus relaxes. The time evolution into different bound states has been discussed in Ref. \onlinecite{Brass:2018}. Similar to the relaxation into bound states, there are several mechanisms for the scattering of electrons into free electronic orbitals. These mechanisms are related to different terms in the Hamiltonian. Electrons can scatter into free electron orbitals due to the changed nuclear potential (Ho changed to Dy). Structures created by this process are often referred to as shake-off structures. Besides the previously mentioned one particle interaction, two particle Coulomb scattering between the remaining electrons of the daughter atom lead to de-excitation via the Auger-Meitner process. One electron from a shallow core state fills the originally created core hole. A second bound electron is simultaneously scattered into a free orbital. The kinetic energy of the unbound electron has continuous eigenvalues. Hence, processes of this kind couple bound-state resonances to the continuous energy spectrum. Formerly sharp excitation peaks get smeared out. As we will show, this leads to asymmetric broadening with large tails strongly affecting the endpoint regime.
	
	To treat these processes and spectral features, we extend the methods developed in Ref. \onlinecite{Brass:2018}. The numerical challange of core level electron capture spectroscopy of a fully interacting atom with 67 electrons is solved using methods from core level x-ray spectroscopy \cite{deGroot:2008wo, Antonides:1977, Zaanen:1986, Tarantelli:1995, Tanaka:1995tl, Bergmann:1999, Rehr:2009eu, Haverkort:2012du, Haverkort:2014hq,Morresi2018}. In section \ref{sec:ECspectrum} we describe the extended method and compare the theoretical predictions to experimental data. Section \ref{sec:LineBroaden} investigates the energy dependence of spectral line-broadening. Here we focus on the underlying scattering channels and cross-sections which shed light on the relaxation processes and explain the modifications of spectral shape due to Auger-Meitner electrons. In section \ref{sec:Implications} we discuss the implications for experiments determining the neutrino masses from electron capture spectra of $^{163}$Ho. Our conclusions can be found in section \ref{sec:Conclusion}. Mathematical and numerical details can be found in the appendices.
	
\section{\label{sec:ECspectrum}The Electron Capture Spectrum}

\begin{figure*}[t]
	\includegraphics[width=\textwidth]{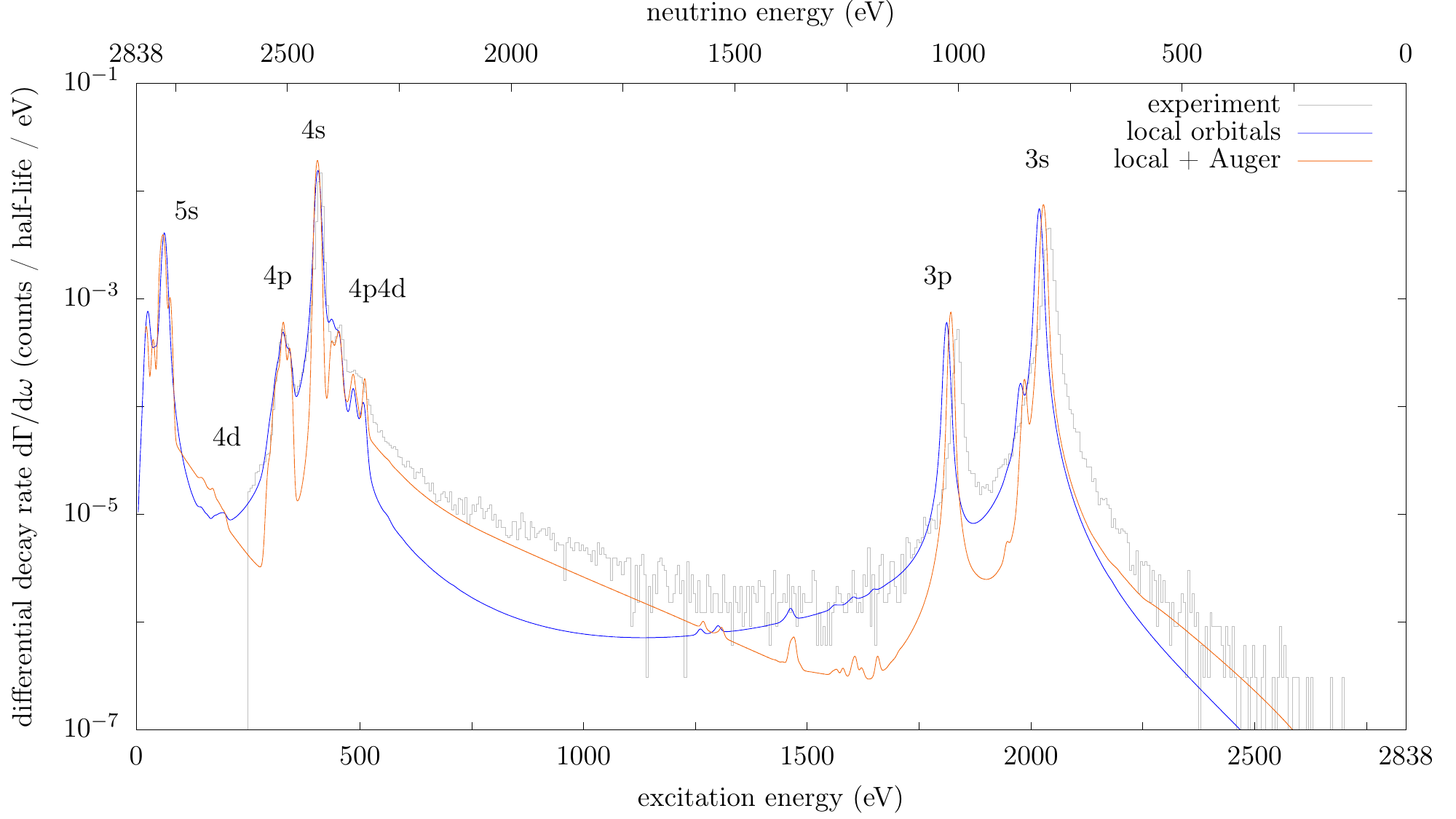}
	\caption{\label{fig:TheoEx} Differential electron capture decay rate $(\mathrm{d}\Gamma / \mathrm{d}\omega)$ per atom per half life as a function of the energy of the neutrino (top scale) or the energy of the electronic excitations (bottom scale) in $^{163}$Ho. The assumed total energy of the decay is $Q=2838$ eV \cite{Eliseev:2015, Clemens:2019}. In grey we plot the experimental spectrum as measured within the ECHO collaboration \cite{Clemens:2019}. In blue we plot the spectrum calculated on a basis of local orbitals artificially broadened to account for life-time not included on the level of theory used in \onlinecite{Brass:2018}. In red we plot the spectrum calculated on a basis including local excitations as well as Auger-Meitner decay into the continuum by solving the Dirac Coulomb equations perturbed by the weak interaction. The theoretical spectra are broadened by a Gaussian of 8 eV to account for the experimental resolution.}
\end{figure*}

	Induced by the weak interaction between nucleons and core-level electrons, $^{163}$Ho can decay via electron capture. Such events create excited $^{163}$Dy daughter atoms with a hole in an atomic orbital from which an electron has been captured. Subsequent de-excitation is mostly driven by Coulomb interactions between the remaining electrons. The resulting de-excitation spectrum involves contributions from states with single and multiple electronic holes. These additional holes are created when electrons are excited to higher bound orbitals or emitted and emerge as Auger-Meitner electrons. Dynamics of electron capture and the inner-atomic de-excitation are described by the Hamiltonian 
\begin{equation}
	\label{eq:Hamiltonian}
	H = H_{\mathrm{Atom}} + K_A + U_A + T \equiv H_0 + T.
\end{equation}
$H_{\text{Atom}}$ contains the kinetic and potential energy of bound electrons as well as their mutual Coulomb interaction in the nuclear potential of a Ho or Dy nucleus. $K_A$ describes the kinetic energy of the released Auger-Meitner electrons and $U_A$ the interaction between bound and Auger-Meitner electrons. Weak interactions are encoded in $T$. It is this operator that is responsible for the electron capture process. Due to the faint nature of weak interaction, $T$ can be treated as a perturbation. $H_{\text{Atom}}$ acts on states with a Ho nucleus ($H_{\text{Ho}}$) before the electron capture event and on states with a Dy nucleus ($H_{\text{Dy}}$) after the capture event. More detail on the different terms in the Hamiltonian can be found in appendices \ref{sec:WaveFuns} and \ref{sec:AugerOperators}. In Ref. \cite{Brass:2018} we showed that the electron capture spectrum can be described within Kubo's formalism using Green's functions
\begin{eqnarray}
	\label{eq:Lehmann}
	\frac{\mathrm{d}\Gamma}{\mathrm{d}\omega}&\propto&\sum_{i=1}^3|U_{ei}|^2\left(Q-\omega\right)\sqrt{\left(Q-\omega\right)^2-m_{\nu_i}^2} \\
	\nonumber &\times&\mathrm{Im}\Big{[}\braket{\psi_{\text{Ho}}|T^{\dagger}\frac{1}{\omega + i 0^+-H_0+E_{\text{Ho}}}T|\psi_{\text{Ho}}}\hspace{0.3cm}\\
	\nonumber &&\quad-\braket{\psi_{\text{Ho}}|T^{\dagger}\frac{1}{\omega + i 0^++H_0-E_{\text{Ho}}}T|\psi_{\text{Ho}}}\Big{]},
\end{eqnarray}
with $Q$ the energy difference between the $^{163}$Ho and $^{163}$Dy ground-state energy, $E_{\Ho}$ the electronic ground state energy of Ho, $\psi_{\Ho}$ the Ho ground-state wave-function, $\omega$ the energy absorbed into electronic excitations during the decay process, $m_{\nu_i}$ the neutrino mass of the $i$-th neutrino mass eigen-state and $U_{ei}$ the Pontecorvo–Maki–Nakagawa–Sakata matrix \cite{Maki62} describing the expansion coefficients of the electron neutrino on the mass eigen-states of the neutrino. Modifications of the spectral shape due to kinetic energy of the created electron neutrino are related to the neutrino phase-space factor:
\begin{equation}
\label{eq:phasefactor}
\sum_{i=1}^3|U_{ei}|^2\left(Q-\omega\right)\sqrt{\left(Q-\omega\right)^2-m_{\nu_i}^2}.
\end{equation}
Owing to conservation of energy, the neutrino's energy is given by the difference of total released energy $Q$ and excitation energy $\omega$ of the daughter atom. The latter is distributed in intensity according to a purely atomic spectrum described by the imaginary part of a Green's function which encodes dynamics of the Ho ground-state $\psi_\Ho$ after an electron capture event due to weak interaction $T$.
	
	In Ref. \cite{Brass:2018} the atomic spectrum was approximated by neglecting Auger-Meitner electrons and the corresponding operators $K_A$ and $U_A$. Thus the Hamiltonian contained a discrete energy spectrum only. When Coulomb interactions couple bound electrons to unbound Auger-Meitner electrons, the former sharp spectral resonances obtain a finite life-time and are consequently broadened. 
	
	The Green's function in the Lehmann representation of the spectrum (eq. \ref{eq:Lehmann}) involves the inverse Hamiltonian projected on the Ho ground-state after electron capture ($T \psi_{Ho}$). This state involves bound electrons solely. We consequently can restrict the inversion on the set of discrete bound-states $\lbrace\psi_b\,|\,H_\Dy\psi_{b}=E_b\psi_b\rbrace$. States including Auger-Meitner electrons contribute via a self-energy
\begin{equation}
\label{eq:Selfenergy}
	\Sigma_{bb'}(\omega) = \braket{\psi_b | U_A\frac{1}{\omega + i0^+ -H_\Dy-K_A+E_\Ho} U_A |\psi_{b'}}.
\end{equation}
This affects the bound-state resonances such that the spectrum becomes
\begin{eqnarray}
\label{eq:SpectrumWithSelfEnergy}
\frac{\mathrm{d}\Gamma}{\mathrm{d}\omega}&\propto&\sum_{i=1}^3|U_{ei}|^2\left(Q-\omega\right)\sqrt{\left(Q-\omega\right)^2-m_{\nu_i}^2} \\
\nonumber &\times&\mathrm{Im}\Big{[}\braket{\psi_{\text{Ho}}|T^{\dagger}\frac{1}{\omega -H_{\text{Dy}}-\Sigma(\omega)+E_{\text{Ho}}}T|\psi_{\text{Ho}}}\hspace{0.3cm}\\
\nonumber &&\quad-\braket{\psi_{\text{Ho}}|T^{\dagger}\frac{1}{\omega +H_{\text{Dy}}+\Sigma(\omega)-E_{\text{Ho}}}T|\psi_{\text{Ho}}}\Big{]}.
\end{eqnarray}
Note that here $H_\Dy$ and $\Sigma(\omega)$ are understood to be projected onto the subspace of bound-states $\lbrace\psi_b\rbrace$ such that they can be expressed as matrices as in eq. \ref{eq:Selfenergy}. For a more rigorous derivation of self-energy and spectral representation see Appendix \ref{sec:DerivationSelfEnergy}.	
	
	The spectrum (eq. \ref{eq:SpectrumWithSelfEnergy}) including bound resonances as well as the Auger-Meitner continuum is plotted in fig. \ref{fig:TheoEx} (red line). The sole parameter that is taken from experiment is the Q-value. The energy positions, line-shapes and widths are a result of solving the Coulomb Dirac equations restricted to a basis set with one free Auger-Meitner electron only. Especially the different broadenings (compare $3s$ and $4s$) as well as the asymmetric line-shapes are calculated \textit{ab initio}. In order to judge the accuracy of our calculations and see the effect of Auger-Meitner decay, we compare the results to the experimental spectrum from Ref. \onlinecite{Clemens:2019} (grey) and to the theoretical calculations from Ref. \onlinecite{Brass:2018} (blue) using bound states only convoluted with a Lorentzian fitted to the experiment. We first of all find that the coupling to the continuum produces much more intensity in the high energy wings of the spectrum than one would expect from a Lorentzian broadening. This asymmetric line broadening has been observed in Ref. \onlinecite{Clemens:2019} where it was attributed to Auger-Meitner decay in combination with the continuum edge onset and auto-ionisation processess. For example, one can reach states just above the $4s$ edge by capturing a $4p$ electron followed by an Auger-Meitner decay transferring a $4s$ electron to the $4p$ shell and emitting a $4f$ valence electron into the vacuum. As these channels open at an energy slightly above the binding energy of the $4s$ shell, one obtains an asymmetric line-shape. Our theoretical calculations find that this is indeed the dominating channell determining the asymmetric line-shape, therefore justifying the interpretation given in Ref. \onlinecite{Clemens:2019}. A second observation is a shift of the energy of the resonances. The $M_1$ edge, i.e. the capture of a $3s$ electron has its maximum at 2040 eV in the experiment, 2019 eV in the theory including only bound states and at 2028 eV in the theory including the self energy due to Auger-Meitner decay. Life-time broadening is normally not associated with energy shifts. Non-the-less, the self energy included in eq. \ref{eq:Selfenergy} due to the Auger-Meitner decay is a fully causal response function describing how these bound states evolve due to the coupling to a continuum by Coulomb interaction. The imaginary part of $\Sigma(\omega)$ which describes the line-broadening is related by the Kramers Kronig relations to its real part. The latter is an energy shift which, as can be seen here, is non-negligible if highly accurate calculations are needed. 

	Even after the inclusion of Auger-Meitner decay, the calculation in fig. \ref{fig:TheoEx} still has some deviations with respect to experiment. Firstly, the experimental resonant energies are shifted by a few eV with respect to theory. Secondly, the theory is sharper than the experimental spectrum. The energy shifts can be related well to the truncation of the one-particle basis functions and the many-body basis states. Scattering of two electrons from an $nl$ orbital into formerly unoccupied orbitals with quantum numbers $n'l'$ can be small, but due to many possible values of $n'$ and $l'$ their total contribution can still be noticeable. Including more orbitals in the basis leads to an exponential scaling of the computation time and quickly becomes intractable. Methods based on renormalisation of the interactions due to these channels within an effective low energy Hilbert space should be tested in the future. The fact that the theoretical spectra are sharper than the experimental spectra indicates that further scattering channels responsible for the decay of the locally bound states are missing. Important candidates for these missing channels are double Auger-Meitner decay and the influence of the environment. In our theory we calculate a single neutral isolated Ho atom. In the experiment the Ho atoms are embedded in gold. The valence electrons of the gold can scatter into the conduction bands opening up additional scattering channels not yet included in the calculation. 

\section{\label{sec:LineBroaden}Energy Dependent Line-Broadening}
	The agreement between theory and experiment in fig. \ref{fig:TheoEx} shows that eq. (\ref{eq:SpectrumWithSelfEnergy}) describes the energy dependent line-broadening quite well. In this section we study the involved processes and give a physical understanding of why the spectrum takes the observed shape. Two effects are dominant. The first is a consequence of energy dependent matrix elements corresponding to Auger-Meitner scattering. This can be fully understood within a single scattering channel by studying single-particle wave-functions of unbound electrons. It is discussed in subsection \ref{subsec:SingleChannel}. The second is known as Fano's effect, where coupling to a continuum of states induces mixing and interference between the resonances. This emerges solely in a multi-channel picture, which is described in subsection \ref{subsec:MultiChannel}.

\subsection{\label{subsec:SingleChannel}Energy-dependent Auger-Meitner Matrix Elements and Free Electron Density of States}

\begin{figure}[t]
	\includegraphics[width=\columnwidth]{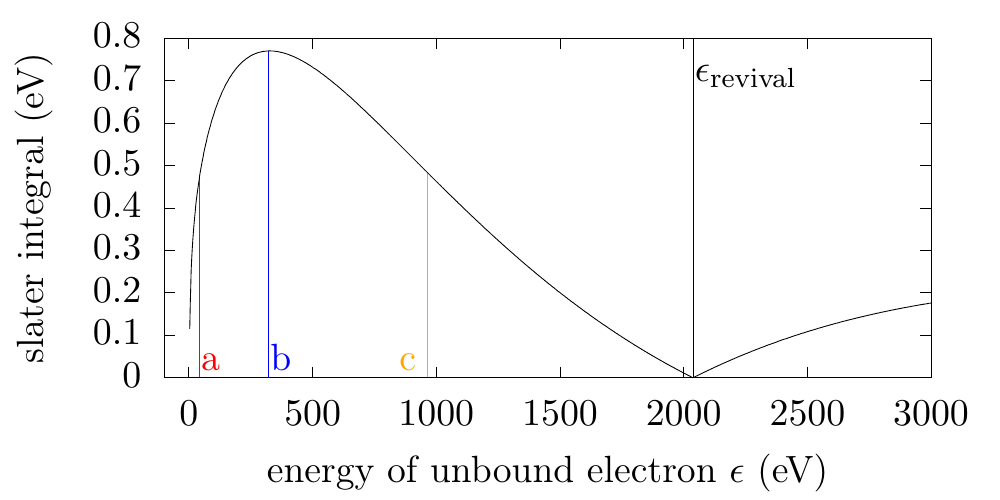} \\
	\includegraphics[width=\columnwidth]{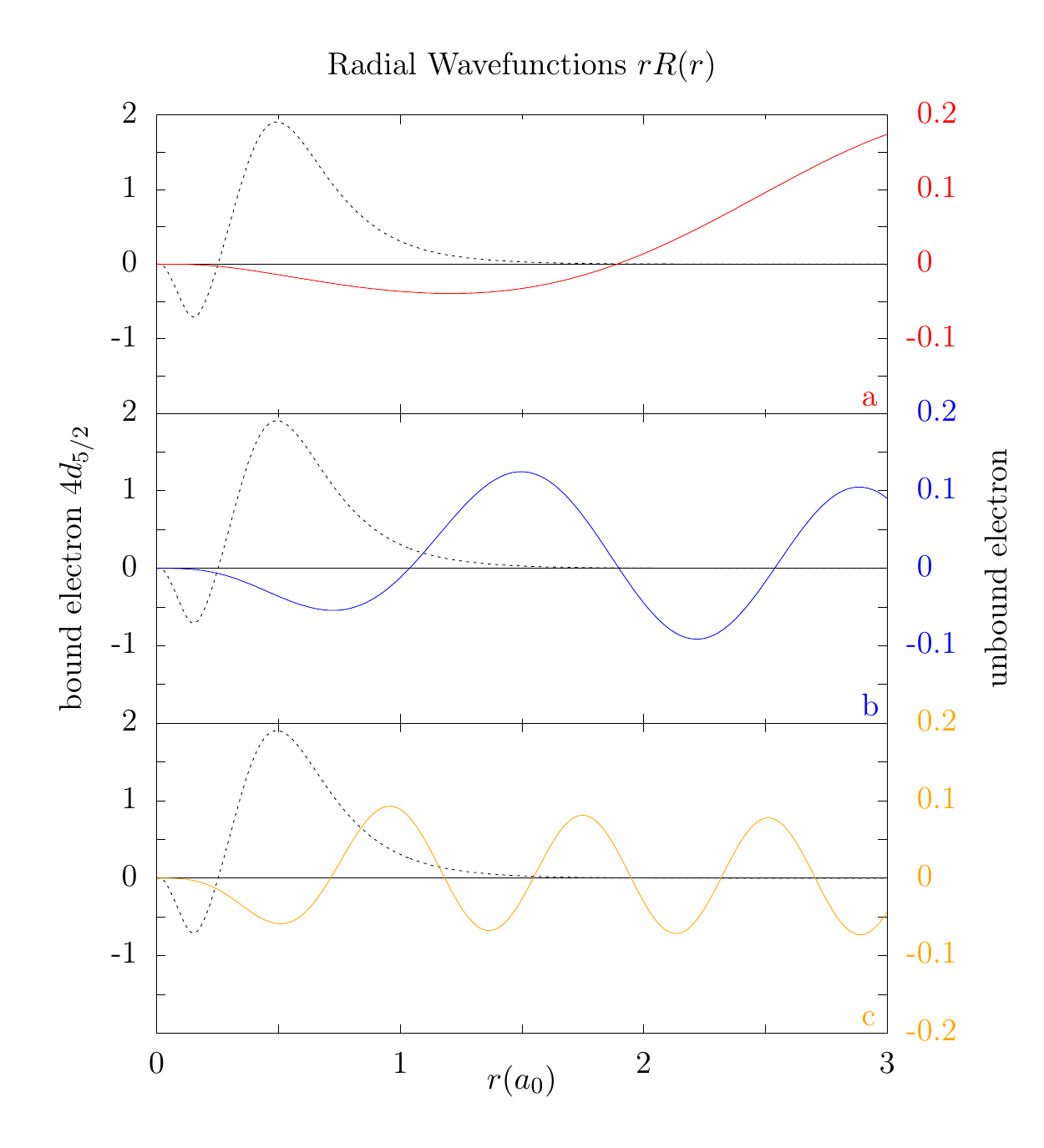} \\
	\caption{\label{fig:WaveFuns}The upper plot shows Slater integrals corresponding to the  $[\Ho ]\underline{4s}\rightarrow [\Ho ]\underline{4p}\underline{4d}+e^{-}_{\epsilon f}$ scattering as function of the single-particle energy $\epsilon$ of the released Auger-Meitner electron with angular momentum $f$. The line at $\epsilon_\mathrm{revival}$ marks the single-particle Auger-Meitner-energy at which the cross-section for Auger-Meitner decay is exactly zero. The colored lines (a-c) indicate the Slater integrals and single-particle energies of the corresponding wave-functions in the lower plot. Here radial wave-functions of unbound (Auger-Meitner) electrons are compared to the bound $4d$ orbital wave-function. Integrals over products of these wave-functions (multiplied by a polynomial in $r$) determine the corresponding Slater integrals shown in the upper plot (color coded).}
\end{figure}

If one looks at the spectral shape of the differential electron capture decay rate one finds that these spectra resemble x-ray core level photo electron spectra (XPES). At the edges one finds an enhanced electron capture decay rate as here the electronic states are at resonance. Above an edge this decay rate is higher than below the edge, because additional channels for electronic excitations are possible. Excess energy can be transferred to an emitted, free electron via the Auger-Meitner process. Hereby the Dy atom is left in an ionized state. These final states are similar to those reached in a photo electron emission event, thereby explaining the similarity between the XPES and EC line-shape. If one simplifies the free electron density of states as well as the Coulomb matrix elements coupling the bound-states to states with an Auger-Meitner electron, one can derive the resulting line-shape to be Mahan like \cite{Mahan:1976, Clemens:2019}. As we want a highly accurate description of the EC spectral shape, one should not make these approximations, but include the detailes of the decay channels in full complexity. The free electron density of states, as well as the Auger-Meitner Coulomb integrals can be calculated such that one can derive the EC line-shape without these approximations. This leads to eq. \ref{eq:SpectrumWithSelfEnergy}, where the broadening now is given by a state dependent self energy (eq. \ref{eq:Selfenergy}). 

In order to understand how this results in the line-shape given in fig. \ref{fig:TheoEx} we here focus on a single decay channel. We define $\psi_b$ to be a single bound excited state of Dy, reached after an electron capture event. This state couples to the continuum via eq. \ref{eq:Selfenergy}, defining the self energy $\Sigma_{bb}(\omega)$. The resulting differential electron capture decay rate for state $\psi_b$ becomes:
\begin{equation}
\frac{\mathrm{d}\Gamma_b}{\mathrm{d}\omega} \propto \mathrm{Im} \left( \frac{1}{\omega-E_b -\Sigma_{bb}(\omega)} - \frac{1}{\omega+E_b +\Sigma_{bb}(\omega)} \right),
\end{equation}
with $E_b$ the excitation energy of the state $\psi_b$ with respect to the Ho ground-state. According to Fermi's golden rule, the life-time of resonance $\psi_b$ is proportional to the inverse transition rate at the resonance energy $\tau_b^{-1}\propto -\mathrm{Im}\left[\Sigma_{bb}(E_b)\right]$. Hence, life-time and energy-dependent broadening are encoded in the diagonal elements of the self-energy. Furthermore, following eq. \ref{eq:Selfenergy} the diagonal parts of the self energy depend on the operator $U_A$ which includes the Coulomb matrix elements between free Auger-Meitner electrons and electrons in orbitals describing the bound-state $\psi_b$.

	To study the impact of those Coulomb matrix elements we focus on a specific case. As an example we take a bound state that belongs to the $4s$ edge of the EC-spectrum. We denote this state by $\ket{[\Ho]\underline{4s}}$ where the underscore implies that there is a core hole in the $4s$ orbital due to electron capture. Note that this is a multi-slater-determinant state, but we will be concerned with this technicality later. This state can decay via many different channels. We focus on the channel where a $4p$ electron fills the hole and transfers the energy-difference to a $4d$ electron which leaves the atom with a kinetic energy $\epsilon$ and angular momentum $l=3$
\begin{equation}
	\label{eq:AMChannel}
	[\Ho ]\underline{4s}\rightarrow [\Ho ]\underline{4p}\underline{4d}+e^{-}_{\epsilon f}
\end{equation}

The coupling strength of this decay is determined by its corresponding Coulomb-Slater integral. This integral depends on the energy of the free electron ($\epsilon$). The size of this integral for the specific case of our example ($4p$ and $4d$ scatter to $4s$ and $\epsilon_f$) is shown in the top plot of fig. \ref{fig:WaveFuns} as a function of the Auger-Meitner electron's kinetic energy. One can observe a very distinct energy dependence, with a clear maximum at 322 eV and a zero at 2037 eV. We can understand this energy dependence by looking at the wave-functions involved. The Coulomb-Slater integrals include the product of the bound radial wave-function ($4d$ orbital) and the Auger-Meitner electron wave function. The lower plot of fig. \ref{fig:WaveFuns} shows, that a low energy, unbound electron has a very small amplitude (red line) in the vicinity of bound $4d$ electrons (dashed line). This leads to a small matrix element for Auger-Meitner decay at low kinetic energy. With increasing energy and decreasing wavelength of the free electron the amplitude of its wave-function in the region of the bound orbitals increases (blue line). With this also the size of the Coulomb matrix elements increases. At a kinetic energy of the free electron of 322 eV, channel (\ref{eq:AMChannel}) has a maximal scattering amplitude. For even higher kinetic energies of the free electron and corresponding smaller wave-lengths (orange line), oscillations of the free electron wave-function in the spatial region of the bound electron reduce the Coulomb-Slater integrals. At 2037 eV this integral becomes zero. Beyond this point the oscillatory behavior of the Auger-Meitner electron's wave-function increases the Slater integral's value again, until the next minimum at even higher kinetic energy and smaller wave-lengths is reached (not shown).
	
\begin{figure}[t]
	\includegraphics[width=\columnwidth]{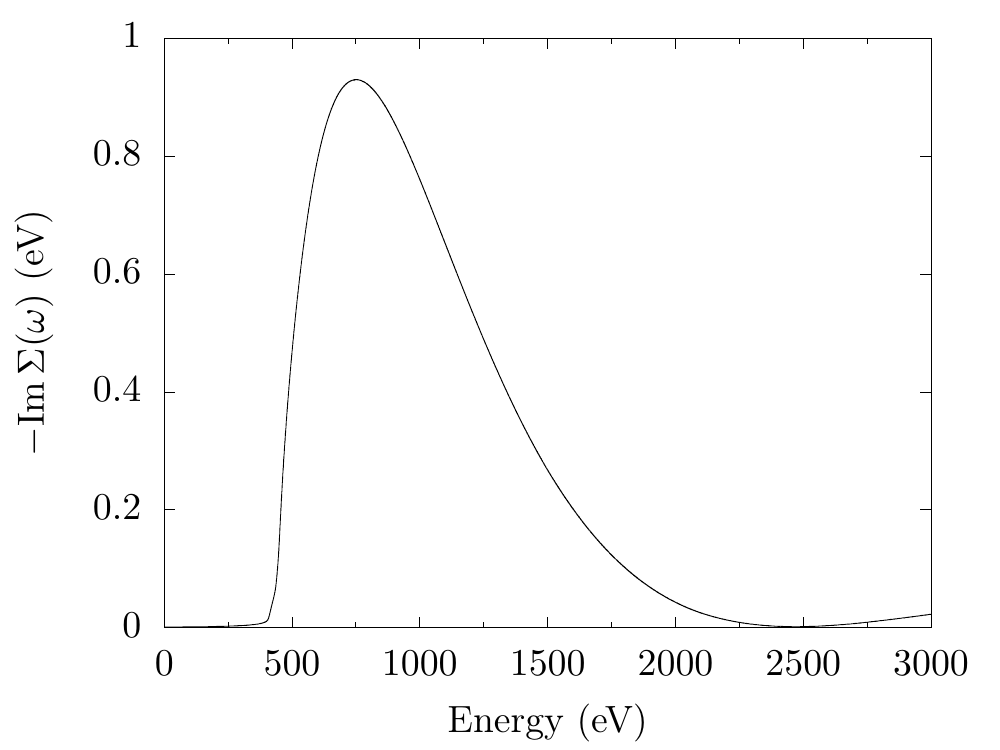} 
	\caption{\label{fig:CrosssectionAuger}Self-energy of the $4s$ core hole due to simultaneous Coulomb scattering of a $4p$ electron into the $4s$ hole and of a $4d$ electron into a free electron orbital with $f$ angular momentum (eq. \ref{eq:AMChannel}). The plot depicts the function defined in eq. \ref{eq:Selfenergy} which is calculated using the Coulomb-Slater integrals shown in fig. \ref{fig:WaveFuns}. }
\end{figure}	
	
	Once the Auger-Meitner energy dependent Coulomb-Slater matrix elements are determined, one can calculate the self energy of state $\ket{\psi_b}=\ket{[\Ho ]\underline{4s}}$ for the single channel used in our example. The operator $U_A$ in eq. \ref{eq:Selfenergy} couples state $\psi_b$ to states with one additional Auger-Meitner electron which can have all possible kinetic energies. As these states are independent of each other, one can sum them separately and obtain the full self energy; i.e. the cross section of Auger-Meitner scattering starting from the state $[\Ho ]\underline{4s}$ as shown in fig. \ref{fig:CrosssectionAuger}. At energies below 400 eV the self energy is small, as there is not enough energy to generate free electron states combined with a $4p$ core hole. Above 400 eV the $4s$ resonance couples to a continuum of Auger-Meitner-states due to Coulomb interaction. The self energy has a maximum around 750 eV, slowly decays and becomes zero around 2500 eV. This is directly related to the Slater integral given in fig. \ref{fig:WaveFuns} which shows, as a function of the kinetic energy of the free electron state, a similar behaviour, albeit shifted by the energy of the $4p$ core hole. 
	
\begin{figure}[t]
	\includegraphics[width=\columnwidth]{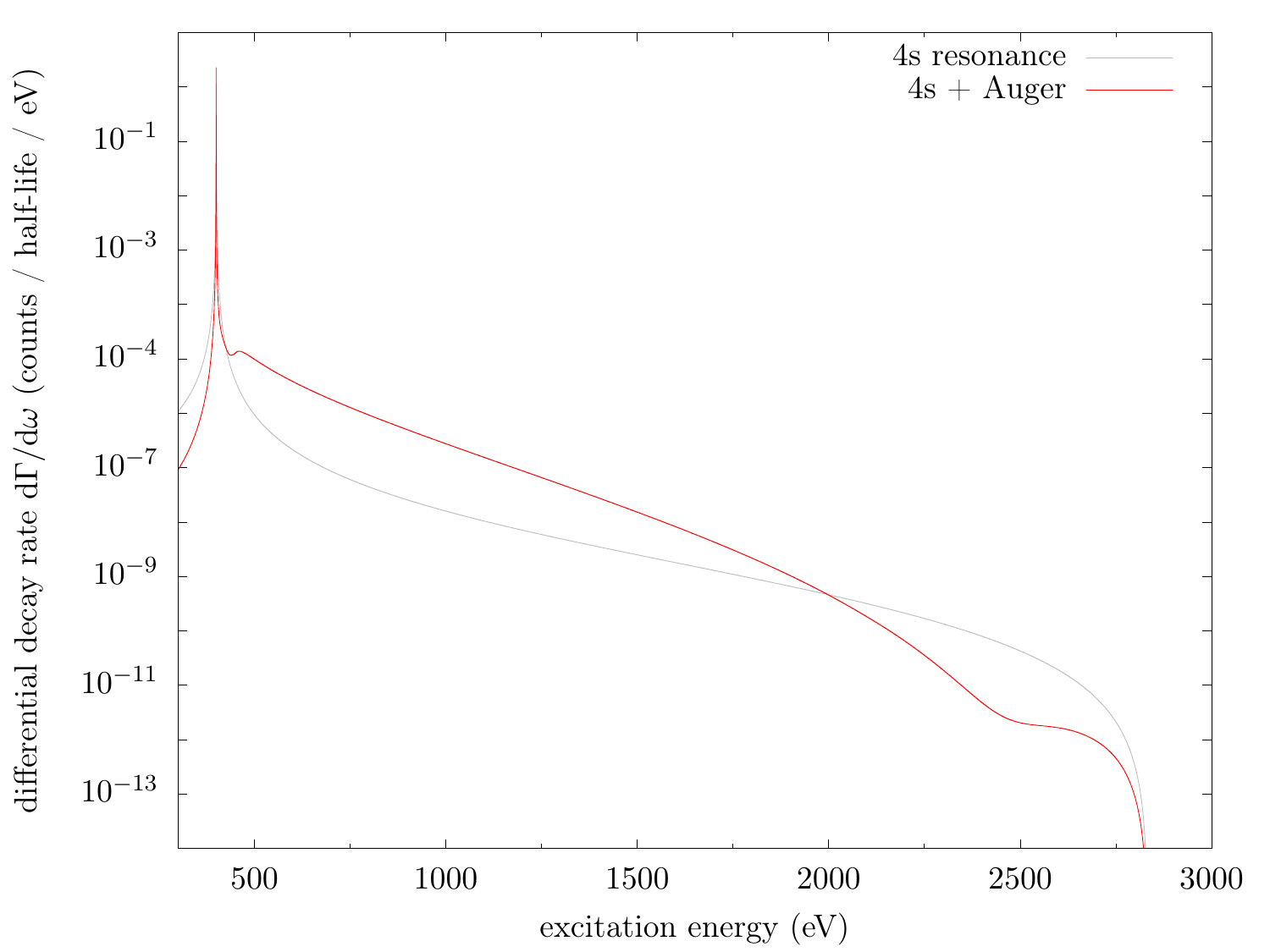} 
	\caption{\label{fig:SingleChannelSpec}Coupling of $4s$ resonance to the continuum of Auger-Meitner electrons via the scattering of a $4p$ electron into the $4s$ orbital while simultaneously scattering a $4p$ electron into a free electron state with $f$ angular momentum. Without Auger-Meitner electrons the resonance would be a single Lorentzian (grey). Additional decay into a continuum results in a wing with increased intensity (red). The wing's shape is determined by the energy dependence of the self energy shown in fig. \ref{fig:CrosssectionAuger}.}
\end{figure}	

With the self energy of this specific state for the given single decay channel one can now calculate the differential decay rate. Fig. \ref{fig:SingleChannelSpec} shows in red the resulting spectrum. In grey the spectrum represented by a single Lorentzian is shown. One can clearly see the asymmetric line-shape of the resonance, which is a consequence of the energy-dependent Coulomb-Slater matrix elements and the resulting energy dependent self energy. The specific energy-dependence of the self energy in fig. \ref{fig:CrosssectionAuger} induces not only the asymmetric broadening, but also the small bump to the right of the peak, as well as the increased spectral weight in the resonance's wing. Even the second rise in self energy above 2500 eV modifies the spectrum, although this effect is suppressed by the neutrino phase-space factor.
	
	In order to calculate the spectral shape as shown in fig. \ref{fig:TheoEx}, one needs to include all possible bound-states $\psi_b$ and all possible decay channels where one bound electron scatters into the created core hole and one bound electron scatters into a free electron state. While the energy dependence of these processes can be understood qualitatively by the overlap between single-particle wave-functions before and after the Auger-Meitner decay, a quantitative treatment needs to include the multi-configurational nature of initial and final states.

\subsection{\label{subsec:MultiChannel}Multi-Channel Auger-Meitner Decay and Fano's Effect}

\begin{figure}[t]
	\includegraphics[width=\columnwidth]{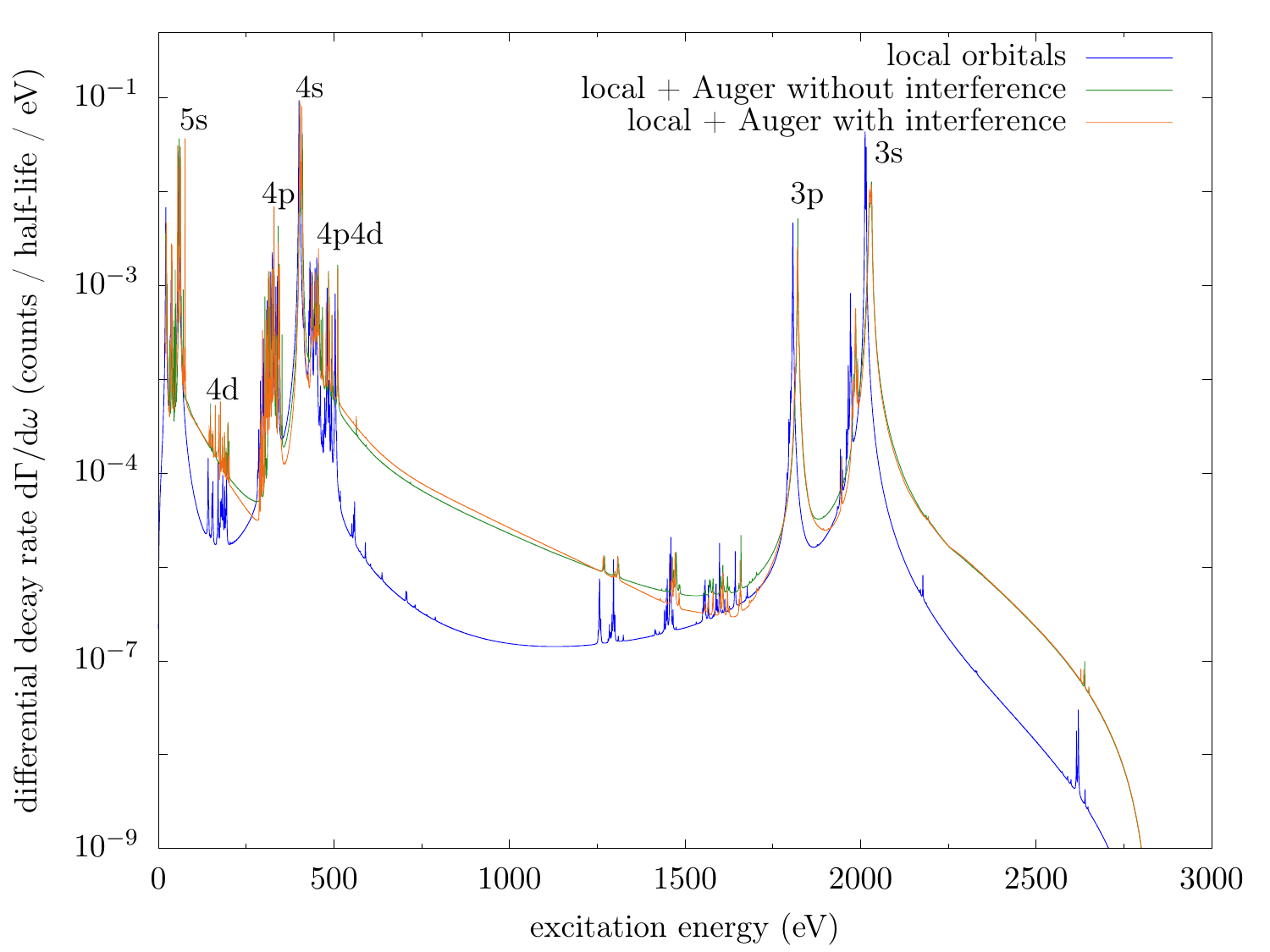} 
	\caption{\label{fig:CompareOldNew}Comparison between the EC spectrum without (blue) \cite{Brass:2018} and with Auger-Meitner decay (red/green). For the green spectrum the self-energy is approximated by its diagonal entries only. The red spectrum also includes the off-diagonal ones. The off diagonal elements of the self energy lead to interference between the resonances resulting in a Fano-like line-shape. }
\end{figure}

	In the previous subsection we demonstrated how energy-dependent Coulomb matrix elements affect the line broadening. We focused on the decay of a single state only, which is encoded in the corresponding diagonal entry of the self-energy as defined in eq. \ref{eq:Selfenergy}. Including multiple scattering channels leads to mixing between bound-state resonances, which in turn is reflected in the off-diagonal elements of the self energy. These describe how different resonances are coupled to the same final states via different scattering channels. For example one can reach a state with a hole in the $4p$ and $4d$ orbital and a free electron by first absorbing a $4s$ electron into the nucleus followed by an Auger-Meitner decay, scattering a $4p$ electron into the $4s$ orbital and emitting a $4d$ electron. One can reach the same state by first absorbing a $3s$ electron into the nucleus followed by an Auger-Meitner decay, scattering a $4p$ electron into the $3s$ orbital and emitting a $4d$ electron. Both pathways lead to the same final state and each pathway comes with its own phase and amplitude. As these are incommensurate, the different channels interfere, leading to Fano-like line-shapes.

In order to determine the interference between different resonances we calculated 226 states $\psi_b$ that represent the bound-state spectrum within the energy range between 0 to 2838 eV with an experimental resolution of 8 eV well. Note that these are not necessarily eigen-states of the Hamiltonian, but linear combinations of states in an energy interval such that the spectrum is sufficiently reproduced. See appendix \ref{sec:Numerics} for more information. In fig. \ref{fig:CompareOldNew} we show in blue the spectrum one obtains with these 226 resonances each convoluted by a single Lorentzian line-shape for later comparison. We also calculated the self-energy for these states due to the Auger-Meitner decay as given in eq. \ref{eq:Selfenergy}. This produces a 226 by 226 matrix. With the use of eq. \ref{eq:SpectrumWithSelfEnergy}, these 226 states $\psi_b$ and the self-energy one can calculate the full spectrum including Auger-Meitner decay and interference between different decay channels. The resulting spectrum is given by the red line in fig. \ref{fig:CompareOldNew}. 

In order to understand the influence of the Fano-effect, i.e. interference between different channels that reach the same final state, we can remove the interference from the full calculation. We then can compare the calculations with and without Fano effect. Removing interference between locally bound states $\psi_b$ that can decay to the same state with an additional Auger-Meitner electron, is done in practice by neglecting the off diagonal elements of the self energy $\Sigma_{bb'}\to\delta_{bb'} \Sigma_{bb}$. The resulting spectrum is given by the green line in fig. \ref{fig:CompareOldNew}. 

If we compare (fig. \ref{fig:CompareOldNew}) the calculation with (red) and without (green) interference between the resonances, we find that the former leads to a more pronounced asymmetry in line-shape. Due to interference, the intensity is reduced left to the resonance and increased right to it. Similar to what was demonstrated in Fano's original paper \cite{Fano:1961}, one observes destructive interference on the low energy tails of resonances and constructive interference on the high energy tails. From fig. \ref{fig:CompareOldNew} it also becomes clear that, with or without Fano interference between the resonances, the line-shape is always rather asymmetric. The red and green line both look very asymmetric compared to the blue line. The most dominant impact on the energy-dependent, asymmetric line-broadening comes from the diagonal elements of the self-energy. The energy dependence of the Auger-Meitner matrix elements, combined with the density of states of the free electron, plays a major role in the determination of the final line-shape.

\subsection{\label{subsec:E0}Ground-state Energy Correction}

Due to the finite number of active orbitals used for the calculation of bound states (see Ref. \onlinecite{Brass:2018} for details) one should expect a substantial error in the total electronic binding energy. As long as the errors for ground-state $\psi_{\mathrm{Ho}}$ and final states $\psi_b$ are the same, this influences the spectrum only marginally. Adding orbitals to the active space in the calculation will lower the energies of both ground-state and excited states in almost equal measure. Thereby the excitation energy is changed only marginally. 

	The inclusion of unbound electrons massively increases the active Hilbert space for the final states. This leads to an energy reduction for all of them and shifts the spectrum to lower energies. These energy-shifts of excited states are of similar magnitude but varying for different resonances. In the calculation energy-shifts are contained in the real part of the self-energy (eq. \ref{eq:SpectrumWithSelfEnergy}), which renormalises the energies of the bound states. As the final states are now treated more accurately than the ground-state, all resonances are shifted to energies which are too low compared to experiment. In order to correct for this, one needs to add the renormalisation of the ground-state energy due to the unbound orbitals. This leads to a shift in ground-state energy by 15.76 eV and improves the agreement between calculated and experimental energies of the resonances.
	
\section{\label{sec:Implications}Implications for experiments obtaining the neutrino mass from $^{163}$Ho}

\begin{figure*}[t]
	\includegraphics[width=\textwidth]{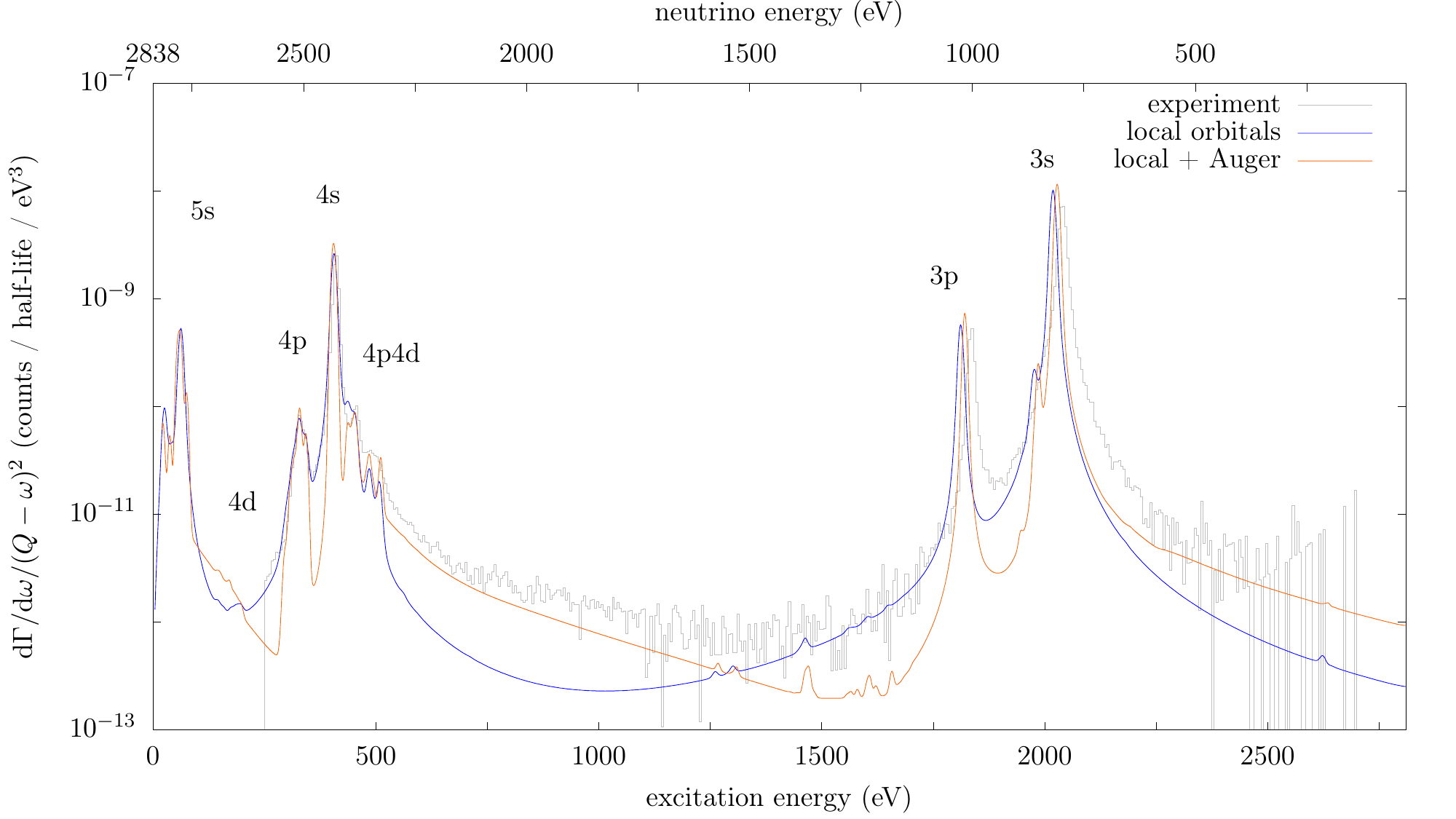}
	\caption{\label{fig:TheoExNoQ} Similar plot as fig \ref{fig:TheoEx} but now with the differential electron capture decay rate divided by the neutrino phase space factor $(Q-\omega)^2$. Scaled differential decay rate as a function of the energy of the neutrino (top scale) or the energy of the electronic excitations (bottom scale) in $^{163}$Ho. The assumed total energy of the decay is $Q=2838$ eV \cite{Eliseev:2015, Clemens:2019}. In grey we plot the experimental spectrum as measured within the ECHO collaboration \cite{Clemens:2019}. In blue we plot the spectrum calculated on a basis of local orbitals artificially broadened to account for life-time not included on the level of theory used in \onlinecite{Brass:2018}. In red we plot the spectrum calculated on a basis including local excitations as well as Auger-Meitner decay into the continuum by solving the Dirac Coulomb equations perturbed by the weak interaction. The theoretical spectra are broadened by a Gaussian of 8 eV to account for the experimental resolution.}
\end{figure*}

The theoretical line-shape of the the differential electron capture nuclear decay rate in $^{163}$Ho has been calculated before using different theoretical levels of complexity \cite{Faessler:2015dg, Faessler:2015ck, Robertson:2015dg, DeRujula:2016cp, Faessler:2017hq, Gastaldo:2017ch,Brass:2018}. It was concluded by Robertson \cite{Robertson:2015dg} that with the level of theory presented at that time "the spectrum shape is not understood well enough to permit a sensitive determination of the neutrino mass in this way". We can now ask the question if with the inclusion of the atomic multiples and Auger decay into bound-states \cite{Brass:2018} as well as the asymmetric line-broadening that follows from our \textit{ab-initio} calculations, one has enough understanding of the spectrum shape at the end-point region to determine the neutrino mass from such experiments. For this it is useful to look at the theoretical spectra without the neutrino phase-space factor (Eq. \ref{eq:phasefactor}).

In fig. \ref{fig:TheoExNoQ} we show the theoretical and experimental spectra divided by the neutrino phase-space factor. We will focus on these spectra at an electronic excitation energy around the $Q$ value of $^{163}$Ho. In this range the neutrino energy is small and here the spectra are most sensitive to the neutrino masses. For the spectra on a local basis, assuming a traditionally used Lorentzian line-shape broadening (blue curve in fig. \ref{fig:TheoExNoQ}) the spectra at the end-point show a curvature making the evaluation of the change of line-shape due to finite neutrino masses hard. For the experimental data and theoretical calculations that include the Auger decay this is different. On a logarithmic plot the scaled differential nuclear decay rate becomes to a very high degree of accuracy linear as a function of decay energy in the region of the $Q$ value of $^{163}$Ho. Despite an overlap of many different involved processes the resulting line-shape at the high energy tail, is surprisingly simple and exponential. This allows for a simple extraction of the neutrino masses. After multiplying the spectrum with $(Q-\omega)^2$ the tail will be linear on a logarithmic scale. The deviation close to $Q$ from a straight line is then related to the neutrino masses. With such a simple line-shape, extracting the three neutrino masses from experimental data with good enough statistics and resolution seems very well possible. 

\section{\label{sec:Conclusion}Conclusion}

	We present an \textit{ab initio} method to calculate the contribution of Auger-Meitner decay to the differential electron capture nuclear decay rate in $^{163}$Ho. Beyond the multiplets and bound resonant structures found for an electron capture spectrum of atomic $^{163}$Ho \cite{Brass:2018}, we observe distinct, asymmetric, energy and state dependent line broadening. The self-energy due to the emission of an electron into the continuum is determined for the many-body propagator on a basis of locally bound states. The imaginary part of the self energy yields the line width, or the lifetime. Its real-part corresponds to an energy shift. The calculation of the self energy includes both the energy dependent Auger-Meitner Coulomb matrix elements, as well as the free electron density of states in the presence of the Ho ion. For the decay of a single state this leads to a Mahan-like, asymmetric line-shape. Besides the sharp state, often referred to as white line, an edge-jump-like feature appears. The Fano effect emerges from interference between different states, which is induced by the self energy coupling different bound many-body states. This reduces the intensity left to the resonance and increases it right to the resonance. 

    The resulting spectrum is in good agreement with recent experimental data obtained by the ECHo collaboration \cite{Clemens:2019}. The experimentally observed increase of spectral weight in the endpoint regime is reproduced by our calculation. Hence, this additional intensity can be understood as a result of atomic relaxation due to Auger-Meitner scattering. On one hand, the increased intensity near the endpoint is good for the experimental determination of the neutrino mass from such spectra, as it leads to higher statistics. On the other hand, the more involved line-shape strengthens the necessity of an accurate theoretical description. We here provide such a theory and show that the differential nuclear decay rate of $^{163}$Ho at the $Q$ value shows a particularly simple behaviour proportional to the neutrino phase-space factor multiplied by an exponential decaying function. This makes the extraction of the neutrino masses from such experiments very feasible, once spectra with enough statistics and high resolution are available.
	
	To improve agreement with experiment further, the effect of the chemical environment should be included. For Ho in Au this will yield an additional line broadening due to low energy excitations of electrons from the valence band of the host into the conduction band. Other possible improvements involve the inclusion of double Auger-Meitner decay states or decay via emission of X-rays. At the same time, one does not need perfect agreement between theory and experiment to be able to deduce the neutrino mass from such measurements. We currently show that with current resolution and statistics all features observed in the spectrum \cite{Clemens:2019} are understood and accounted for. 
	
	With the availability of high quality data emerging due to recent developments in metallic magnetic calorimeters \cite{Enss:2005ue} it will be interesting to apply this method  to other elements whose isotopes show nuclear decay via electron capture \cite{Alotiby:2018,Koehler:2018,Croce:2016,Loidl:2018,Inoyatov:2007,Inoyatov:2011,Inoyatov:2012,Inoyatov:2013,Inoyatov:2015,Tee:2019,Casey:1968,Porter:1974,Vetter,Voytas:2002,Fontanelli:1996,Meunier,Hartmann:1985,Coron:2012}.
	
\section{Acknowledgments}

This project has received funding from the EMPIR programme co-financed by the Participating States and from the European Union’s Horizon 2020 research and innovation programme within the 17FUN02 MetroMMC project. Part of this research was funded by the Deutsche Forschungsgemeinschaft (DFG, German Research Foundation) - 400329440, Research Unit FOR2202 Neutrino Mass Determination by Electron Capture in 163Ho, ECHo (funding under Grant No. HA6108/2-1).  

\clearpage

\appendix

\section*{Appendix}

	Our calculation of the energy dependent line-broadening due to Auger-Meitner scattering is implemented as an extension in \textsc{Quanty}, a many-body script language \cite{Haverkort:2014hq, Haverkort:2016hz, QuantyWebsite}. The appendices are devoted to a description of the methods used to perform these calculations. In appendix \ref{sec:WaveFuns} we describe the construction of free electron one particle wave-functions. Any complete basis set is possible to use, but some are more convenient than others. These one particle wave-functions are needed to calculate the matrix elements of the Coulomb interaction $U_{\text{A}}$ between bound and unbound electrons, which is done in appendix \ref{sec:AugerOperators}. In appendix \ref{sec:DerivationSelfEnergy} we derive the expressions for self-energy (eq. \ref{eq:Selfenergy}) and EC spectrum (eq. \ref{eq:SpectrumWithSelfEnergy}). Appendix \ref{sec:Numerics} treats numerical aspects of the calculation.
	
\section{\label{sec:WaveFuns}Auger-Meitner Electron Single Particle Wave-Functions}

	The calculation of the Coulomb matrix elements (eq. \ref{eq:AugerUMatrixElement}) which couple bound and unbound electrons requires single particle wave-functions of the Auger-Meitner electrons. As the wave-functions of unbound states are not square integrable we use a countable, orthonormal basis of square integrable wave-packets (WP). We follow the methods developed by Rubtsova, Kukulin and Pomerantsev \cite{Rubtsova:2015}. These WPs are constructed from the spinors $\varsigma_{qjlm}(r,\theta,\varphi)$ which are eigen-functions of the free Dirac-Hamiltonian in spherical coordinates. We discretized the continuous energy spectrum by dividing the spectrum of radial momentum $q$ into intervals. We take the values for momenta such that we obtain a set of electrons with equidistant energy spacing $\lbrace\epsilon_n=c\sqrt{q_n^2 + m^2c^2}\,\, |\,\, n=1...\infty\rbrace $. We take an energy spacing of 2 eV between successive $\epsilon_n$. This yields a resonable energy resolution for the self energy. The set of free electron basis functions is truncated at an energy of 4 keV. This yields 2000 wave-packets per angular momentum to describe the Auger-Meitner electrons. The cut-off scale at 4 keV is chosen to be larger than the Q-value in order to avoid artificial side-effects due to a truncated basis set. The wave-packets are  obtained as
	\begin{equation}
	\label{eq:WavePackets}
	\varphi_{\epsilon_njlm}(r,\theta ,\varphi) = \int_{q_n}^{q_n+\delta q}\!\varsigma_{qjlm}(r,\theta ,\varphi)\mathrm{d}q
	\end{equation}
	These functions span the space of unbound single-particle states, but are not orthogonal to the bound-state wave-functions. The final basis $\lbrace\phi_{\epsilon_njlm}(\mathbf{r})\rbrace$ of unbound states is constructed by orthonormalizing the $\varphi_{\epsilon_njlm}(\mathbf{r})$ with respect to the bound-state wavefunctions $\psi_{njlm}(\mathbf{r})$. The latter are obtained from the density functional theory program FPLO \cite{Koepernik:1999uw, Opahle:1999tx, Eschrig:2004wn} as  discussed in \cite{Brass:2018}.
	
	This basis allows for a construction of the second quantized operators described in the next section.

\section{\label{sec:AugerOperators}Auger-Meitner Electron Operators}
	With the discretization of the continuous energy spectrum from appendix \ref{sec:WaveFuns} and the corresponding WPs as basis functions, we can construct the kinetic energy operator for Auger-Meitner electrons as
\begin{equation}
	\label{eq:AugerKinetic}
	K_\mathrm{A} = \sum_{\epsilon_n}\sum_{l=0}^5\sum_{j=l-\frac{1}{2}}^{l+\frac{1}{2}}\sum_{m=-j}^j\epsilon_n c^\dagger_{\epsilon_n l j m}c_{\epsilon_n l j m}
\end{equation}
	where $c^\dagger_{\epsilon_n l j m}/c_{\epsilon_n l j m}$ creates/annihilates an electron in the unbound state $\phi_{\epsilon_njlm}(\mathbf{r})$ with kinetic energy $\epsilon_n$.  In the following we will denote the four quantum-numbers in the above equation collectively by $\epsilon\equiv\lbrace\epsilon_n ,l,j,m\rbrace$ and a sum over $\epsilon$ is understood as four sums over all quantum-numbers.
	
	The knowledge of the Auger-Meitner electron wave-functions $\phi_{\epsilon}(\mathbf{r})$ enables us further to calculate the matrix elements for Coulomb scattering of bound electrons into unbound states via
\begin{equation}
	\label{eq:AugerUMatrixElement}
	U_{\epsilon abc} = -\int\!\frac{\left(\phi_\epsilon^*\left(\mathbf{r}_1\right)\cdot\psi_{b}\left(\mathbf{r}_1\right)\right)\left(\psi_{a}^*\left(\mathbf{r}_2\right)\cdot\psi_{c}\left(\mathbf{r}_2\right)\right)}{|\mathbf{r}_1-\mathbf{r}_2|}\mathrm{d}\mathbf{r}_1\mathrm{d}\mathbf{r}_2
\end{equation}
	where Roman indices denote quantum numbers $\lbrace n,j,l,m\rbrace$ of bound state wavefunctions $\psi_{njlm}(\mathbf{r})$. The operator that governs all Auger-Meitner scatterings is given by
\begin{equation}
	\label{eq:AugerU}
	U_\mathrm{A} =\sum_\epsilon\sum_{ijk}U_{\epsilon ijk}c^\dagger_{\epsilon}c^\dagger_{i}c_{k}c_{j}+\mathrm{h.c.}\equiv \sum_\epsilon U_\epsilon
\end{equation}
	The full dynamics of the remaining electrons after EC is described by the Hamiltonian
\begin{equation}
	\label{eq:FullHamiltonian}
	H = H_\Dy +K_\mathrm{A} +U_\mathrm{A}\equiv H_0+\sum_\epsilon U_\epsilon
\end{equation}
	Here, interactions between Auger-Meitner electrons are neglected, since we assume that in most of the relaxation processes there is only a single Auger-Meitner electron present at a time.

\section{\label{sec:DerivationSelfEnergy}Derivation of the Self-Energy}

	With the wave-functions for Auger-Meitner electrons in appendix \ref{sec:WaveFuns} and their second quantized operators in appendix \ref{sec:AugerOperators} we now derive eq. \ref{eq:Selfenergy} for the self-energy.
	
	If one includes Auger-Meitner scattering, calculating the EC spectrum (eq. \ref{eq:Lehmann}) involves the inversion of the Hamiltonian (eq. \ref{eq:FullHamiltonian}). Here, the interaction $U_\epsilon$ (eq. \ref{eq:AugerU}) creates or annihilates one Auger-Meitner electron with energy $\epsilon$. The operator only couples sub-spaces of Fock-space that differ in the number of Auger-Meitner electrons with energy $\epsilon$ by one. After the EC event, there are only bound electrons and one hole. Final states become more unlikely to influence the spectrum the higher the number of Auger-Meitner electrons they contain. To make use of this property, we divide the full Fock-space $\mathcal{F}$ into subspaces with fixed numbers of Auger-Meitner electrons by introducing the following orthogonal projections. $P^{(0)}$ projects onto the sub-space of all configurations that do not involve Auger-Meitner electrons. We denote this subspace by $P^{(0)}\mathcal{F}\equiv\lbrace P^{(0)}\psi\, |\,\psi\in\mathcal{F}\rbrace$. For every Auger-Meitner energy $\epsilon$ the operator $P^{(1)}_\epsilon$ projects onto the sub-space of configurations that involve exactly one Auger-Meitner electron with energy $\epsilon$. Onto the sub-space with two Auger-Meitner electrons with energies $\epsilon_1$ and $\epsilon_2$ we can project with $P^{(2)}_{\epsilon_1,\epsilon_2}$ and so on. By construction the decomposition of $\mathcal{F}$ and the completeness relation read
\begin{equation}
	\label{eq:Completeness}
	\mathcal{F}=\bigcup_{n=0}^N\left(\bigcup_{\epsilon_1,...,\epsilon_n}P^{(n)}_{\epsilon_1,...,\epsilon_n}\mathcal{F}\right)\hspace{0.5cm}\mathbb{1} = \sum_{n=0}^{N}\sum_{\epsilon_1,...,\epsilon_n}P^{(n)}_{\epsilon_1,...,\epsilon_n}
\end{equation}
	where $N$ is the total number of electrons in the system. Since Auger-Meitner energies have been discretized and cut-off at 4 keV, the Fock-space is finite dimensional and we can express the Hamiltonian as a block-matrix by using the projection operators. We restrict us on the subspace that includes configurations with at most one Auger-Meitner electron
\begin{widetext}
\begin{equation}
	\label{eq:HamiltonianMatrix}
 	H|_{\left(P^{(0)}\mathcal{F}\right)\bigcup\,\left(\bigcup_\epsilon P^{(1)}_\epsilon\mathcal{F}\right)}=\left(\begin{array}{cccccc}
 		P^{(0)}H_0P^{(0)} & P^{(0)}U_{\epsilon_1}P^{(1)}_{\epsilon_1} & P^{(0)}U_{\epsilon_2}P^{(1)}_{\epsilon_2} & P^{(0)}U_{\epsilon_3}P^{(1)}_{\epsilon_3}& \cdots & P^{(0)}U_{\epsilon_n}P^{(1)}_{\epsilon_n}\\
 		P^{(1)}_{\epsilon_1}U_{\epsilon_1} P^{(0)} & P^{(1)}_{\epsilon_1} H_0P^{(1)}_{\epsilon_1}  & 0 & 0 & \cdots&0 \\
 		P^{(1)}_{\epsilon_2}U_{\epsilon_2}P^{(0)} & 0 &P^{(1)}_{\epsilon_2}H_0P^{(1)}_{\epsilon_2} & 0& \cdots & 0 \\
 		P^{(1)}_{\epsilon_3}U_{\epsilon_3}P^{(0)} & 0 & 0&P^{(1)}_{\epsilon_3}H_0P^{(1)}_{\epsilon_3} &\cdots &0 \\
 		\vdots &\vdots & \vdots &0 &\ddots &\vdots\\
 		P^{(1)}_{\epsilon_n}U_{\epsilon_n}P^{(0)} & 0 & 0& 0 &\cdots & P^{(1)}_{\epsilon_n}H_0P^{(1)}_{\epsilon_n} \\
 	\end{array}\right)
\end{equation} 
\end{widetext}
	From the Lehmann representation of the spectrum (eq. \ref{eq:Lehmann}) we see that the calculation only involves $P^{(0)}(z-H)^{-1}P^{(0)}$ which we can directly read off from (\ref{eq:HamiltonianMatrix})
\begin{equation}
	\label{eq:PHP}
	P^{(0)}(z-H)^{-1}P^{(0)}=\left(P^{(0)}z-P^{(0)}H_0P^{(0)}-\Sigma(z)\right)^{-1}
\end{equation}
where
\begin{equation}
	\Sigma(z)=\sum_\epsilon\left[ P^{(0)}U_\epsilon P^{(1)}_\epsilon\left(z-H_0\right)^{-1}P^{(1)}_\epsilon U_\epsilon P^{(0)}\right]
\end{equation}
	This expression can be further simplified by recognizing that $ P^{(0)}U_\epsilon P^{(1)}_{\epsilon'}=0$ if $\epsilon\neq\epsilon'$, since $U_\epsilon$ annihilates (creates) an Auger-Meitner electron with energy $\epsilon$. Thus, using the completeness relation (eq. \ref{eq:Completeness}) one gets $P^{(0)}U_\epsilon\mathbb{1}= P^{(0)}U_\epsilon P^{(1)}_{\epsilon}$ such that the self-energy is reduced to
\begin{equation}
	\label{eq:SigmaMatrix}
	\Sigma(z)=P^{(0)}\sum_\epsilon \left[ U_\epsilon \left(z-H_0\right)^{-1} U_\epsilon \right]P^{(0)}
\end{equation}
	The great advantage of this equation is that it involves only the projection onto the space without Auger-Meitner electrons which is a large reduction of active space. If we use the Krylov-basis of bound multi-configuration states $\lbrace\psi_{b}\rbrace$ introduced in section \ref{sec:ECspectrum} to express the projection $P^{(0)}=\sum_{\psi_{b}}\ket{\psi_{b}}\bra{\psi_{b}}$, the self-energy $\Sigma(z)$ becomes equivalent to eq. \ref{eq:Selfenergy}. Taking the expectation value of eq. \ref{eq:PHP} with respect to $T\ket{\psi_{\text{Ho}}}$ one arrives at the final expression for the EC spectrum (eq. \ref{eq:SpectrumWithSelfEnergy}).
	
	In this derivation two approximations have been made. First we neglected the interaction between Auger-Meitner electrons in the construction of the Hamiltonian (eq. \ref{eq:FullHamiltonian}). Second we neglected subspaces with more than one Auger-Meitner electron. The first one is a fundamental simplification, since it determines the form of the Hamiltonian as matrix in eq. \ref{eq:HamiltonianMatrix}. This form is essential for the whole derivation. Further interactions between unbound electrons would change this form, drastically increase the size of Fock-space and couple multiple subspaces making the calculation no longer feasible. The other approximation is not this essential but yields an important performance boost in the calculation. One could extend it to include states with two or more unbound electrons, which would in analogy yield a self-energy for the states with a single unbound electron. However, this already shows that the effect is expected to be small as it involves two Auger-Meitner-decays and hence is of order $O(U_A^2)$.

\section{\label{sec:Numerics}Numerical Calculation of the Self-energy}

In this section we describe how to calculate the expression for the self-energy $\Sigma_{bb'}(\omega)$ derived in appendix \ref{sec:DerivationSelfEnergy} numerically. As a first step we describe how to obtain an optimum basis set $\psi_b$ to represent the spectrum for bound states only. The second step describes the calculation of the self energy on this basis.

In order to calculate the set given by the states $\psi_b$ we start from the many-body Ho ground-state wave-function. We then annihilate one of the $ns$ or $np_{1/2}$ electrons creating 20 new many-electron states. Starting from these states we generate a Krylov basis using a block Lanczos routine. These states represent the spectral function well, but are not necessarily eigen-states of the Hamiltonian. In total we include 2000 states in our Krylov basis. Most of these states are outside the spectral window of interest. In order to reduce the dimension of the self energy we diagonalize the Hamiltonian on a basis of the Krylov states and only keep the states that have an energy below the $Q$ value. In total we kept 226 states. For more detail on the generation of the Krylov basis and the Block-Lanczos procedure see \onlinecite{Brass:2018}.

Once we have a basis for the bound states that represent the spectral function well, we need to calculate the Auger-Meitner decay of these states. Firstly we need to determine the coupling between these bound states and states with one electron less and an additional Auger-Meitner electron. Given the energy discretization of the unbound electrons from appendix \ref{sec:WaveFuns}, the operator $U_\mathrm{A}$ contains about $10^7$ terms. The number of involved Slater determinants for each of the wave-functions $\psi_{b}$ is of similar magnitude. Hence, it is essential to rewrite the self-energy such that the calculation can be easily parallelized on multiple CPUs and such that the amount of repeated operations is reduced to a minimum. To achieve this, we insert $U_\epsilon$ from eq. \ref{eq:AugerU} into eq. \ref{eq:SigmaMatrix} and make use of $K_\mathrm{A}c^\dagger_\epsilon\ket{\psi_{b}} = \epsilon c^\dagger_\epsilon\ket{\psi_{b}}$ yielding
\begin{widetext}
\begin{eqnarray}
	\label{eq:SEreducedForm}
	\Sigma_{bb'}(\omega)&=&\sum_\epsilon\sum_{ijk}\sum_{i'j'k'}U_{ijk\epsilon}^* \braket{\psi_{b}|c^\dagger_{i}c^\dagger_{j}c_{\epsilon}c_{k} \,\left(\omega+i0^+-H_\Dy-K_\mathrm{A}\right)^{-1}c^\dagger_{\epsilon}c^\dagger_{i'}c_{k'}c_{j'} |\psi_{b'}} U_{\epsilon i'j'k'}\nonumber\\
	&=&\sum_\epsilon\sum_{ijk}\sum_{i'j'k'}U_{ijk\epsilon}^* \braket{\psi_{b}|c^\dagger_{i}c^\dagger_{j}c_{k} \,\left(\omega+i0^+-H_\Dy-\epsilon\right)^{-1}c^\dagger_{i'}c_{k'}c_{j'} |\psi_{b'}} U_{\epsilon i'j'k'}\nonumber\\
	&\equiv&\sum_{\epsilon}\sum_{\tau\tau'}U^*_{\tau\epsilon}\braket{\tau|R_{H_\Dy}(\omega+i0^+-\epsilon)|\tau'}U_{\epsilon\tau'}\hspace{0.5cm}\mathrm{with}\hspace{0.5cm}\ket{\tau =(i,j,k)}=c^\dagger_{k}c_{j}c_{i} \ket{\psi_{b}}
\end{eqnarray}
\end{widetext}
	In the last step we introduced the resolvent of the Dy Hamiltonian
\begin{equation}
	R_{H_\Dy}\left(z\right)=\left(z-H_\Dy\right)^{-1}
\end{equation}	
	The resolvent matrix elements $\braket{\tau|R_{H_\Dy}|\tau'}$ do not depend on the created Auger-Meitner electron and its kinetic energy. Therefore, the resolvent only needs to be calculated once - independent of the unbound electron's energy. The energy dependent line-broadening enters via the slater-integrals $U_{ijk\epsilon}$ which can be calculated fully in parallel. Eq. \ref{eq:SEreducedForm} yields an efficient form to evaluate the self-energy. The most resource consuming task is the evaluation of the matrix elements of the resolvent on the sub-space spanned by the vectors $\ket{\tau}$. This evaluation can be done by a conventional Block-Lanczos routine as implemented in \textsc{Quanty} \cite{Haverkort:2014hq, Haverkort:2016hz, QuantyWebsite}.


\begin{thebibliography}{29}%
\makeatletter
\providecommand \@ifxundefined [1]{%
\@ifx{#1\undefined}
}%
\providecommand \@ifnum [1]{%
\ifnum #1\expandafter \@firstoftwo
\else \expandafter \@secondoftwo
\fi
}%
\providecommand \@ifx [1]{%
\ifx #1\expandafter \@firstoftwo
\else \expandafter \@secondoftwo
\fi
}%
\providecommand \natexlab [1]{#1}%
\providecommand \enquote  [1]{``#1''}%
\providecommand \bibnamefont  [1]{#1}%
\providecommand \bibfnamefont [1]{#1}%
\providecommand \citenamefont [1]{#1}%
\providecommand \href@noop [0]{\@secondoftwo}%
\providecommand \href [0]{\begingroup \@sanitize@url \@href}%
\providecommand \@href[1]{\@@startlink{#1}\@@href}%
\providecommand \@@href[1]{\endgroup#1\@@endlink}%
\providecommand \@sanitize@url [0]{\catcode `\\12\catcode `\$12\catcode
`\&12\catcode `\#12\catcode `\^12\catcode `\_12\catcode `\%12\relax}%
\providecommand \@@startlink[1]{}%
\providecommand \@@endlink[0]{}%
\providecommand \url  [0]{\begingroup\@sanitize@url \@url }%
\providecommand \@url [1]{\endgroup\@href {#1}{\urlprefix }}%
\providecommand \urlprefix  [0]{URL }%
\providecommand \Eprint [0]{\href }%
\providecommand \doibase [0]{http://dx.doi.org/}%
\providecommand \selectlanguage [0]{\@gobble}%
\providecommand \bibinfo  [0]{\@secondoftwo}%
\providecommand \bibfield  [0]{\@secondoftwo}%
\providecommand \translation [1]{[#1]}%
\providecommand \BibitemOpen [0]{}%
\providecommand \bibitemStop [0]{}%
\providecommand \bibitemNoStop [0]{.\EOS\space}%
\providecommand \EOS [0]{\spacefactor3000\relax}%
\providecommand \BibitemShut  [1]{\csname bibitem#1\endcsname}%
\let\auto@bib@innerbib\@empty

\bibitem{Katrin2019}M. Aker \textit{et al}. (KATRIN Collaboration), Phys. Rev. Lett. \textbf{123}, 221802 (2019).

\bibitem{Fermi34}E. Fermi, Zeitschrift F{\"u}r Physik a Hadrons and Nuclei \textbf{88}, 161 (1934).

\bibitem [{\citenamefont {De~R{\'u}jula}\ and\ \citenamefont
	{Lusignoli}(1982)}]{DeRujula:1982us}%
\BibitemOpen
\bibfield  {author} {\bibinfo {author} {\bibfnamefont {A.}~\bibnamefont
		{De~R{\'u}jula}}\ and\ \bibinfo {author} {\bibfnamefont {M.}~\bibnamefont
		{Lusignoli}},\ }\href@noop {} {\bibfield  {journal} {\bibinfo  {journal}
		{Physics Letters B}\ }\textbf {\bibinfo {volume} {118}},\ \bibinfo {pages}
	{429} (\bibinfo {year} {1982})}\BibitemShut {NoStop}%
\bibitem [{\citenamefont {Alpert}\ \emph {et~al.}(2015)\citenamefont {Alpert},
	\citenamefont {Balata}, \citenamefont {Bennett}, \citenamefont {Biasotti},
	\citenamefont {Boragno}, \citenamefont {Brofferio}, \citenamefont {Ceriale},
	\citenamefont {Corsini}, \citenamefont {Day}, \citenamefont {De~Gerone},
	\citenamefont {Dressler}, \citenamefont {Faverzani}, \citenamefont {Ferri},
	\citenamefont {Fowler}, \citenamefont {Gatti}, \citenamefont {Giachero},
	\citenamefont {Hays-Wehle}, \citenamefont {Heinitz}, \citenamefont {Hilton},
	\citenamefont {K{\"o}ster}, \citenamefont {Lusignoli}, \citenamefont {Maino},
	\citenamefont {Mates}, \citenamefont {Nisi}, \citenamefont {Nizzolo},
	\citenamefont {Nucciotti}, \citenamefont {Pessina}, \citenamefont
	{Pizzigoni}, \citenamefont {Puiu}, \citenamefont {Ragazzi}, \citenamefont
	{Reintsema}, \citenamefont {Gomes}, \citenamefont {Schmidt}, \citenamefont
	{Schumann}, \citenamefont {Sisti}, \citenamefont {Swetz}, \citenamefont
	{Terranova},\ and\ \citenamefont {Ullom}}]{Alpert:2015gi}%
\BibitemOpen
\bibfield  {author} {\bibinfo {author} {\bibfnamefont {B.}~\bibnamefont
		{Alpert}}, \bibinfo {author} {\bibfnamefont {M.}~\bibnamefont {Balata}},
	\bibinfo {author} {\bibfnamefont {D.}~\bibnamefont {Bennett}}, \bibinfo
	{author} {\bibfnamefont {M.}~\bibnamefont {Biasotti}}, \bibinfo {author}
	{\bibfnamefont {C.}~\bibnamefont {Boragno}}, \bibinfo {author} {\bibfnamefont
		{C.}~\bibnamefont {Brofferio}}, \bibinfo {author} {\bibfnamefont
		{V.}~\bibnamefont {Ceriale}}, \bibinfo {author} {\bibfnamefont
		{D.}~\bibnamefont {Corsini}}, \bibinfo {author} {\bibfnamefont {P.~K.}\
		\bibnamefont {Day}}, \bibinfo {author} {\bibfnamefont {M.}~\bibnamefont
		{De~Gerone}}, \bibinfo {author} {\bibfnamefont {R.}~\bibnamefont {Dressler}},
	\bibinfo {author} {\bibfnamefont {M.}~\bibnamefont {Faverzani}}, \bibinfo
	{author} {\bibfnamefont {E.}~\bibnamefont {Ferri}}, \bibinfo {author}
	{\bibfnamefont {J.}~\bibnamefont {Fowler}}, \bibinfo {author} {\bibfnamefont
		{F.}~\bibnamefont {Gatti}}, \bibinfo {author} {\bibfnamefont
		{A.}~\bibnamefont {Giachero}}, \bibinfo {author} {\bibfnamefont
		{J.}~\bibnamefont {Hays-Wehle}}, \bibinfo {author} {\bibfnamefont
		{S.}~\bibnamefont {Heinitz}}, \bibinfo {author} {\bibfnamefont
		{G.}~\bibnamefont {Hilton}}, \bibinfo {author} {\bibfnamefont
		{U.}~\bibnamefont {K{\"o}ster}}, \bibinfo {author} {\bibfnamefont
		{M.}~\bibnamefont {Lusignoli}}, \bibinfo {author} {\bibfnamefont
		{M.}~\bibnamefont {Maino}}, \bibinfo {author} {\bibfnamefont
		{J.}~\bibnamefont {Mates}}, \bibinfo {author} {\bibfnamefont
		{S.}~\bibnamefont {Nisi}}, \bibinfo {author} {\bibfnamefont {R.}~\bibnamefont
		{Nizzolo}}, \bibinfo {author} {\bibfnamefont {A.}~\bibnamefont {Nucciotti}},
	\bibinfo {author} {\bibfnamefont {G.}~\bibnamefont {Pessina}}, \bibinfo
	{author} {\bibfnamefont {G.}~\bibnamefont {Pizzigoni}}, \bibinfo {author}
	{\bibfnamefont {A.}~\bibnamefont {Puiu}}, \bibinfo {author} {\bibfnamefont
		{S.}~\bibnamefont {Ragazzi}}, \bibinfo {author} {\bibfnamefont
		{C.}~\bibnamefont {Reintsema}}, \bibinfo {author} {\bibfnamefont {M.~R.}\
		\bibnamefont {Gomes}}, \bibinfo {author} {\bibfnamefont {D.}~\bibnamefont
		{Schmidt}}, \bibinfo {author} {\bibfnamefont {D.}~\bibnamefont {Schumann}},
	\bibinfo {author} {\bibfnamefont {M.}~\bibnamefont {Sisti}}, \bibinfo
	{author} {\bibfnamefont {D.}~\bibnamefont {Swetz}}, \bibinfo {author}
	{\bibfnamefont {F.}~\bibnamefont {Terranova}}, \ and\ \bibinfo {author}
	{\bibfnamefont {J.}~\bibnamefont {Ullom}},\ }\href@noop {} {\bibfield
	{journal} {\bibinfo  {journal} {Eur. Phys. J. C}\ }\textbf {\bibinfo {volume}
		{75}},\ \bibinfo {pages} {27} (\bibinfo {year} {2015})}\BibitemShut {NoStop}%
\bibitem [{\citenamefont {Croce}\ \emph {et~al.}(2016)\citenamefont {Croce},
	\citenamefont {Hoover}, \citenamefont {Rabin}, \citenamefont {Bond},
	\citenamefont {Wolfsberg}, \citenamefont {Schmidt},\ and\ \citenamefont
	{Ullom}}]{Croce:2016dp}%
\BibitemOpen
\bibfield  {author} {\bibinfo {author} {\bibfnamefont {M.~P.}\ \bibnamefont
		{Croce}}, \bibinfo {author} {\bibfnamefont {A.~S.}\ \bibnamefont {Hoover}},
	\bibinfo {author} {\bibfnamefont {M.~W.}\ \bibnamefont {Rabin}}, \bibinfo
	{author} {\bibfnamefont {E.~M.}\ \bibnamefont {Bond}}, \bibinfo {author}
	{\bibfnamefont {L.~E.}\ \bibnamefont {Wolfsberg}}, \bibinfo {author}
	{\bibfnamefont {D.~R.}\ \bibnamefont {Schmidt}}, \ and\ \bibinfo {author}
	{\bibfnamefont {J.~N.}\ \bibnamefont {Ullom}},\ }\href@noop {} {\bibfield
	{journal} {\bibinfo  {journal} {Journal of Low Temperature Physics}\ }\textbf
	{\bibinfo {volume} {184}},\ \bibinfo {pages} {938} (\bibinfo {year}
	{2016})}\BibitemShut {NoStop}%
\bibitem [{\citenamefont {Gastaldo}\ \emph {et~al.}(2017)\citenamefont
	{Gastaldo}, \citenamefont {Blaum}, \citenamefont {Chrysalidis}, \citenamefont
	{Day~Goodacre}, \citenamefont {Domula}, \citenamefont {Door}, \citenamefont
	{Dorrer}, \citenamefont {D~llmann}, \citenamefont {Eberhardt}, \citenamefont
	{Eliseev}, \citenamefont {Enss}, \citenamefont {Faessler}, \citenamefont
	{Filianin}, \citenamefont {Fleischmann}, \citenamefont {Fonnesu},
	\citenamefont {Gamer}, \citenamefont {Haas}, \citenamefont {Hassel},
	\citenamefont {Hengstler}, \citenamefont {Jochum}, \citenamefont {Johnston},
	\citenamefont {Kebschull}, \citenamefont {Kempf}, \citenamefont {Kieck},
	\citenamefont {K~ster}, \citenamefont {Lahiri}, \citenamefont {Maiti},
	\citenamefont {Mantegazzini}, \citenamefont {Marsh}, \citenamefont
	{Neroutsos}, \citenamefont {Novikov}, \citenamefont {Ranitzsch},
	\citenamefont {Rothe}, \citenamefont {Rischka}, \citenamefont {Saenz},
	\citenamefont {Sander}, \citenamefont {Schneider}, \citenamefont {Scholl},
	\citenamefont {Sch~ssler}, \citenamefont {Schweiger}, \citenamefont {{\v
			S}imkovic}, \citenamefont {Stora}, \citenamefont {Sz~cs}, \citenamefont
	{T~rler}, \citenamefont {Veinhard}, \citenamefont {Weber}, \citenamefont
	{Wegner}, \citenamefont {Wendt},\ and\ \citenamefont
	{Zuber}}]{Gastaldo:2017ch}%
\BibitemOpen
\bibfield  {author} {\bibinfo {author} {\bibfnamefont {L.}~\bibnamefont
		{Gastaldo}}, \bibinfo {author} {\bibfnamefont {K.}~\bibnamefont {Blaum}},
	\bibinfo {author} {\bibfnamefont {K.}~\bibnamefont {Chrysalidis}}, \bibinfo
	{author} {\bibfnamefont {T.}~\bibnamefont {Day~Goodacre}}, \bibinfo {author}
	{\bibfnamefont {A.}~\bibnamefont {Domula}}, \bibinfo {author} {\bibfnamefont
		{M.}~\bibnamefont {Door}}, \bibinfo {author} {\bibfnamefont {H.}~\bibnamefont
		{Dorrer}}, \bibinfo {author} {\bibfnamefont {C.~E.}\ \bibnamefont
		{D~llmann}}, \bibinfo {author} {\bibfnamefont {K.}~\bibnamefont {Eberhardt}},
	\bibinfo {author} {\bibfnamefont {S.}~\bibnamefont {Eliseev}}, \bibinfo
	{author} {\bibfnamefont {C.}~\bibnamefont {Enss}}, \bibinfo {author}
	{\bibfnamefont {A.}~\bibnamefont {Faessler}}, \bibinfo {author}
	{\bibfnamefont {P.}~\bibnamefont {Filianin}}, \bibinfo {author}
	{\bibfnamefont {A.}~\bibnamefont {Fleischmann}}, \bibinfo {author}
	{\bibfnamefont {D.}~\bibnamefont {Fonnesu}}, \bibinfo {author} {\bibfnamefont
		{L.}~\bibnamefont {Gamer}}, \bibinfo {author} {\bibfnamefont
		{R.}~\bibnamefont {Haas}}, \bibinfo {author} {\bibfnamefont {C.}~\bibnamefont
		{Hassel}}, \bibinfo {author} {\bibfnamefont {D.}~\bibnamefont {Hengstler}},
	\bibinfo {author} {\bibfnamefont {J.}~\bibnamefont {Jochum}}, \bibinfo
	{author} {\bibfnamefont {K.}~\bibnamefont {Johnston}}, \bibinfo {author}
	{\bibfnamefont {U.}~\bibnamefont {Kebschull}}, \bibinfo {author}
	{\bibfnamefont {S.}~\bibnamefont {Kempf}}, \bibinfo {author} {\bibfnamefont
		{T.}~\bibnamefont {Kieck}}, \bibinfo {author} {\bibfnamefont
		{U.}~\bibnamefont {K~ster}}, \bibinfo {author} {\bibfnamefont
		{S.}~\bibnamefont {Lahiri}}, \bibinfo {author} {\bibfnamefont
		{M.}~\bibnamefont {Maiti}}, \bibinfo {author} {\bibfnamefont
		{F.}~\bibnamefont {Mantegazzini}}, \bibinfo {author} {\bibfnamefont
		{B.}~\bibnamefont {Marsh}}, \bibinfo {author} {\bibfnamefont
		{P.}~\bibnamefont {Neroutsos}}, \bibinfo {author} {\bibfnamefont {Y.~N.}\
		\bibnamefont {Novikov}}, \bibinfo {author} {\bibfnamefont {P.~C.-O.}\
		\bibnamefont {Ranitzsch}}, \bibinfo {author} {\bibfnamefont {S.}~\bibnamefont
		{Rothe}}, \bibinfo {author} {\bibfnamefont {A.}~\bibnamefont {Rischka}},
	\bibinfo {author} {\bibfnamefont {A.}~\bibnamefont {Saenz}}, \bibinfo
	{author} {\bibfnamefont {O.}~\bibnamefont {Sander}}, \bibinfo {author}
	{\bibfnamefont {F.}~\bibnamefont {Schneider}}, \bibinfo {author}
	{\bibfnamefont {S.}~\bibnamefont {Scholl}}, \bibinfo {author} {\bibfnamefont
		{R.~X.}\ \bibnamefont {Sch~ssler}}, \bibinfo {author} {\bibfnamefont
		{C.}~\bibnamefont {Schweiger}}, \bibinfo {author} {\bibfnamefont
		{F.}~\bibnamefont {{\v S}imkovic}}, \bibinfo {author} {\bibfnamefont
		{T.}~\bibnamefont {Stora}}, \bibinfo {author} {\bibfnamefont
		{Z.}~\bibnamefont {Sz~cs}}, \bibinfo {author} {\bibfnamefont
		{A.}~\bibnamefont {T~rler}}, \bibinfo {author} {\bibfnamefont
		{M.}~\bibnamefont {Veinhard}}, \bibinfo {author} {\bibfnamefont
		{M.}~\bibnamefont {Weber}}, \bibinfo {author} {\bibfnamefont
		{M.}~\bibnamefont {Wegner}}, \bibinfo {author} {\bibfnamefont
		{K.}~\bibnamefont {Wendt}}, \ and\ \bibinfo {author} {\bibfnamefont
		{K.}~\bibnamefont {Zuber}},\ }\href@noop {} {\bibfield  {journal} {\bibinfo
		{journal} {Eur. Phys. J. Spec. Top.}\ }\textbf {\bibinfo {volume} {226}},\
	\bibinfo {pages} {1623} (\bibinfo {year} {2017})}\BibitemShut {NoStop}%
\bibitem [{\citenamefont {Faessler}\ \emph {et~al.}(2015)\citenamefont
	{Faessler}, \citenamefont {Enss}, \citenamefont {Gastaldo},\ and\
	\citenamefont {{\v S}imkovic}}]{Faessler:2015dg}%
\BibitemOpen
\bibfield  {author} {\bibinfo {author} {\bibfnamefont {A.}~\bibnamefont
		{Faessler}}, \bibinfo {author} {\bibfnamefont {C.}~\bibnamefont {Enss}},
	\bibinfo {author} {\bibfnamefont {L.}~\bibnamefont {Gastaldo}}, \ and\
	\bibinfo {author} {\bibfnamefont {F.}~\bibnamefont {{\v S}imkovic}},\
}\href@noop {} {\bibfield  {journal} {\bibinfo  {journal} {Phys. Rev. C}\
	}\textbf {\bibinfo {volume} {91}},\ \bibinfo {pages} {064302} (\bibinfo
	{year} {2015})}\BibitemShut {NoStop}%
\bibitem [{\citenamefont {Faessler}\ and\ \citenamefont {{\v
			S}imkovic}(2015)}]{Faessler:2015ck}%
\BibitemOpen
\bibfield  {author} {\bibinfo {author} {\bibfnamefont {A.}~\bibnamefont
		{Faessler}}\ and\ \bibinfo {author} {\bibfnamefont {F.}~\bibnamefont {{\v
				S}imkovic}},\ }\href@noop {} {\bibfield  {journal} {\bibinfo  {journal}
		{Phys. Rev. C}\ }\textbf {\bibinfo {volume} {91}},\ \bibinfo
	{pages} {045505}  (\bibinfo {year} {2015})}\BibitemShut {NoStop}%
\bibitem [{\citenamefont {Robertson}(2015)}]{Robertson:2015dg}%
\BibitemOpen
\bibfield  {author} {\bibinfo {author} {\bibfnamefont {R.~G.~H.}\
		\bibnamefont {Robertson}},\ }\href@noop {} {\bibfield  {journal} {\bibinfo
		{journal} {Phys. Rev. C}\ }\textbf {\bibinfo {volume} {91}},\ \bibinfo
	{pages} {035504} (\bibinfo {year} {2015})}\BibitemShut {NoStop}%
\bibitem [{\citenamefont {De~R{\'u}jula}\ and\ \citenamefont
	{Lusignoli}(2016)}]{DeRujula:2016cp}%
\BibitemOpen
\bibfield  {author} {\bibinfo {author} {\bibfnamefont {A.}~\bibnamefont
		{De~R{\'u}jula}}\ and\ \bibinfo {author} {\bibfnamefont {M.}~\bibnamefont
		{Lusignoli}},\ }\href@noop {} {\bibfield  {journal} {\bibinfo  {journal} {J.
			High Energ. Phys.}\ }\textbf {\bibinfo {volume} {2016}},\ \bibinfo {pages}
	{15} (\bibinfo {year} {2016})}\BibitemShut {NoStop}%
\bibitem [{\citenamefont {Faessler}\ \emph {et~al.}(2017)\citenamefont
	{Faessler}, \citenamefont {Gastaldo},\ and\ \citenamefont {{\v
			S}imkovic}}]{Faessler:2017hq}%
\BibitemOpen
\bibfield  {author} {\bibinfo {author} {\bibfnamefont {A.}~\bibnamefont
		{Faessler}}, \bibinfo {author} {\bibfnamefont {L.}~\bibnamefont {Gastaldo}},
	\ and\ \bibinfo {author} {\bibfnamefont {F.}~\bibnamefont {{\v S}imkovic}},\
}\href@noop {} {\bibfield  {journal} {\bibinfo  {journal} {Phys. Rev. C}\
	}\textbf {\bibinfo {volume} {95}},\ \bibinfo {pages} {045502} (\bibinfo
	{year} {2017})}\BibitemShut {NoStop}%
	\bibitem{Ramitzsch17} P. C.-O. Ranitzsch, C. Hassel, M. Wegner, D. Hengstler, S. Kempf, A. Fleischmann, C. Enss, L. Gastaldo, A. Herlert, and K. Johnston, Phys. Rev. Lett. \textbf{119}, 122501 (2017).
\bibitem [{\citenamefont {Bra\ss}\ \emph {et~al.}(2018)\citenamefont
{Bra\ss}, \citenamefont {Enss}, \citenamefont {Gastaldo},\citenamefont {Green},\ and\
\citenamefont {Haverkort}}]{Brass:2018}%
\BibitemOpen
\bibfield  {author} {\bibinfo {author} {\bibfnamefont {M.}~\bibnamefont
{Bra\ss}}, \bibinfo {author} {\bibfnamefont {C.}~\bibnamefont {Enss}},
\bibinfo {author} {\bibfnamefont {L.}~\bibnamefont {Gastaldo}},
\bibinfo {author} {\bibfnamefont {R. J.}~\bibnamefont {Green}},\ and\
\bibinfo {author} {\bibfnamefont {M. W.}~\bibnamefont {Haverkort}},\
}\href@noop {} {\bibfield  {journal} {\bibinfo  {journal} {Phys. Rev. C}\ }\textbf {\bibinfo {volume} {97}},\ \bibinfo {pages} {054620}
(\bibinfo {year} {2018})}\BibitemShut {NoStop}%
\bibitem{Clemens:2019} C. Velte, F. Ahrens, A. Barth, K. Blaum, M. Bra{\ss}, M. Door, H. Dorrer, Ch. E. D{\"u}llmann, S. Eliseev, C. Enss, P. Filianin, A. Fleischmann, L. Gastaldo, A. Goeggelmann, T. Day Goodacre, M. W. Haverkort, D. Hengstler, J. Jochum, K. Johnston, M. Keller, S. Kempf, T. Kieck, C. M. K{\"o}nig, U. K{\"o}ster, K. Kromer, F. Mantegazzini, B. Marsh, Yu. N. Novikov, F. Piquemal, C. Riccio, D. Richter, A. Rischka, S. Rothe, R. X. Sch{\"u}ssler, Ch. Schweiger, T. Stora, M. Wegner, K. Wendt, M. Zampaolo, and K. Zuber, Eur. Phys. J. C  \textbf{79}, 1026 (2019).
\bibitem [{\citenamefont {de~Groot}\ and\ \citenamefont
	{Kotani}(2008)}]{deGroot:2008wo}%
\BibitemOpen
\bibfield  {author} {\bibinfo {author} {\bibfnamefont {F.~M.~F.}\
		\bibnamefont {de~Groot}}\ and\ \bibinfo {author} {\bibfnamefont
		{A.}~\bibnamefont {Kotani}},\ }\href@noop {} {\emph {\bibinfo {title} {{Core
				Level Spectroscopy of Solids}}}}\ (\bibinfo  {publisher} {CRC Press},\
\bibinfo {year} {2008})\BibitemShut {NoStop}%
\bibitem{Antonides:1977}E. Antonides, E. C. Janse, and G. A. Sawatzky, Phys. Rev. B \textbf{15}, 1669 (1977).
\bibitem{Zaanen:1986}J. Zaanen and G. A. Sawatzky, Phys. Rev. B \textbf{33}, 8074 (1986).
\bibitem [{\citenamefont {Tanaka}\ and\ \citenamefont
	{Jo}(1995)}]{Tanaka:1995tl}%
\BibitemOpen
\bibfield  {author} {\bibinfo {author} {\bibfnamefont {A.}~\bibnamefont
		{Tanaka}}\ and\ \bibinfo {author} {\bibfnamefont {T.}~\bibnamefont {Jo}},\
}\href@noop {} {\bibfield  {journal} {\bibinfo  {journal} {J. Phys. Soc.
			Jpn.}\ } (\bibinfo {year} {1995})}\BibitemShut {NoStop}%
\bibitem{Tarantelli:1995} F. Tarantelli, L.S. Cederbaum and A. Sgamellotti, Journal of Electron Spectroscopy and Related Phenomena \textbf{76}, 47 (1995).
\bibitem{Bergmann:1999}U. Bergmann, P. Glatzel, F. M. F. de Groot, and S. P. Cramer, J. Am. Chem. Soc. \textbf{121}, 4926 (1999).
\bibitem [{\citenamefont {Rehr}\ \emph {et~al.}(2009)\citenamefont {Rehr},
	\citenamefont {Kas}, \citenamefont {Prange}, \citenamefont {Sorini},
	\citenamefont {Takimoto},\ and\ \citenamefont {Vila}}]{Rehr:2009eu}%
\BibitemOpen
\bibfield  {author} {\bibinfo {author} {\bibfnamefont {J.~J.}\ \bibnamefont
		{Rehr}}, \bibinfo {author} {\bibfnamefont {J.~J.}\ \bibnamefont {Kas}},
	\bibinfo {author} {\bibfnamefont {M.~P.}\ \bibnamefont {Prange}}, \bibinfo
	{author} {\bibfnamefont {A.~P.}\ \bibnamefont {Sorini}}, \bibinfo {author}
	{\bibfnamefont {Y.}~\bibnamefont {Takimoto}}, \ and\ \bibinfo {author}
	{\bibfnamefont {F.}~\bibnamefont {Vila}},\ }\href@noop {} {\bibfield
	{journal} {\bibinfo  {journal} {C. R. Physique}\ }\textbf {\bibinfo {volume}
		{10}},\ \bibinfo {pages} {548} (\bibinfo {year} {2009})}\BibitemShut
{NoStop}%
\bibitem [{\citenamefont {Haverkort}\ \emph {et~al.}(2012)\citenamefont
	{Haverkort}, \citenamefont {Zwierzycki},\ and\ \citenamefont
	{Andersen}}]{Haverkort:2012du}%
\BibitemOpen
\bibfield  {author} {\bibinfo {author} {\bibfnamefont {M.~W.}\ \bibnamefont
		{Haverkort}}, \bibinfo {author} {\bibfnamefont {M.}~\bibnamefont
		{Zwierzycki}}, \ and\ \bibinfo {author} {\bibfnamefont {O.~K.}\ \bibnamefont
		{Andersen}},\ }\href@noop {} {\bibfield  {journal} {\bibinfo  {journal}
		{Phys. Rev. B}\ }\textbf {\bibinfo {volume} {85}},\ \bibinfo {pages} {165113}
	(\bibinfo {year} {2012})}\BibitemShut {NoStop}%
\bibitem [{\citenamefont {Haverkort}\ \emph {et~al.}(2014)\citenamefont
	{Haverkort}, \citenamefont {Sangiovanni}, \citenamefont {Hansmann},
	\citenamefont {Toschi}, \citenamefont {Lu},\ and\ \citenamefont
	{Macke}}]{Haverkort:2014hq}%
\BibitemOpen
\bibfield  {author} {\bibinfo {author} {\bibfnamefont {M.~W.}\ \bibnamefont
		{Haverkort}}, \bibinfo {author} {\bibfnamefont {G.}~\bibnamefont
		{Sangiovanni}}, \bibinfo {author} {\bibfnamefont {P.}~\bibnamefont
		{Hansmann}}, \bibinfo {author} {\bibfnamefont {A.}~\bibnamefont {Toschi}},
	\bibinfo {author} {\bibfnamefont {Y.}~\bibnamefont {Lu}}, \ and\ \bibinfo
	{author} {\bibfnamefont {S.}~\bibnamefont {Macke}},\ }\href@noop {}
{\bibfield  {journal} {\bibinfo  {journal} {Europhys. Lett.}\ }\textbf
	{\bibinfo {volume} {108}},\ \bibinfo {pages} {57004} (\bibinfo {year}
	{2014})}\BibitemShut {NoStop}%
\bibitem{Morresi2018} Tommaso Morresi, Simone Taioli, and Stefano Simonucci, Adv. Theory Simul. \textbf{1} 1800086 (2018).
\bibitem{Maki62}Z. Maki, M. Nakagawa, and S. Sakata, Progress of Theoretical Physics \textbf{28}, 870 (1962).
\bibitem{Eliseev:2015}S. Eliseev, K. Blaum, M. Block, S. Chenmarev, H. Dorrer, C. E. D{\"u}llmann, C. Enss, P. E. Filianin, L. Gastaldo, M. Goncharov, U. K{\"o}ster, F. Lautenschl{\"a}ger, Y. N. Novikov, A. Rischka, R. X. Sch{\"u}ssler, L. Schweikhard, and A. T{\"u}rler, Phys. Rev. Lett. \textbf{115}, 062501 (2015).
\bibitem{Mahan:1976}G. D. Mahan, Phys. Rev. \textbf{163}, 612 (1967).
\bibitem [{\citenamefont {Fano}(1961)}]{Fano:1961}%
\BibitemOpen
\bibfield  {author} {\bibinfo {author} {\bibfnamefont {U.}\ \bibnamefont
		{Fano}},\ }\href@noop {} {\bibfield  {journal} {\bibinfo  {journal} {Phys. Rev.}\ }\textbf {\bibinfo {volume} {124}},\ \bibinfo {pages} {1866} (\bibinfo {year} {1961})}\BibitemShut {NoStop}%
\bibitem [{\citenamefont {Haverkort}(2016)}]{Haverkort:2016hz}%
\BibitemOpen
\bibfield  {author} {\bibinfo {author} {\bibfnamefont {M.~W.}\ \bibnamefont
		{Haverkort}},\ }\href@noop {} {\bibfield  {journal} {\bibinfo  {journal} {J.
			Phys.: Conf. Ser.}\ }\textbf {\bibinfo {volume} {712}},\ \bibinfo {pages}
	{012001} (\bibinfo {year} {2016})}\BibitemShut {NoStop}%
\bibitem{QuantyWebsite}One can download the code and find documentation at http://www.Quanty.org
\bibitem [{\citenamefont {Rubtsova}\ \emph {et~al.}(2015)\citenamefont
{Rubtsova}, \citenamefont {Kukulin}, \citenamefont {Pomerantsev}}]{Rubtsova:2015}%
\BibitemOpen
\bibfield  {author} {\bibinfo {author} {\bibfnamefont O.A.}~\bibnamefont
{Rubtsova}, \bibinfo {author} {\bibfnamefont {V.I.}~\bibnamefont {Kukulin}},\ and\
\bibinfo {author} {\bibfnamefont {V.N.}~\bibnamefont {Pomerantsev}},\
}\href@noop {} {\bibfield  {journal} {\bibinfo  {journal} {Ann. Phys.}\ }\textbf {\bibinfo {volume} {360}},\ \bibinfo {pages} {613}
(\bibinfo {year} {2015})}\BibitemShut {NoStop}%
\bibitem [{\citenamefont {Koepernik}\ and\ \citenamefont
{Eschrig}(1999)}]{Koepernik:1999uw}%
\BibitemOpen
\bibfield  {author} {\bibinfo {author} {\bibfnamefont {K.}~\bibnamefont
{Koepernik}}\ and\ \bibinfo {author} {\bibfnamefont {H.}~\bibnamefont
{Eschrig}},\ }\href@noop {} {\bibfield  {journal} {\bibinfo  {journal} {Phys.
Rev. B}\ }\textbf {\bibinfo {volume} {59}},\ \bibinfo {pages} {1743}
(\bibinfo {year} {1999})}\BibitemShut {NoStop}%
\bibitem [{\citenamefont {Opahle}\ \emph {et~al.}(1999)\citenamefont {Opahle},
\citenamefont {Koepernik},\ and\ \citenamefont {Eschrig}}]{Opahle:1999tx}%
\BibitemOpen
\bibfield  {author} {\bibinfo {author} {\bibfnamefont {I.}~\bibnamefont
{Opahle}}, \bibinfo {author} {\bibfnamefont {K.}~\bibnamefont {Koepernik}}, \
and\ \bibinfo {author} {\bibfnamefont {H.}~\bibnamefont {Eschrig}},\
}\href@noop {} {\bibfield  {journal} {\bibinfo  {journal} {Phys. Rev. B}\
}\textbf {\bibinfo {volume} {60}},\ \bibinfo {pages} {14035} (\bibinfo {year}
{1999})}\BibitemShut {NoStop}%
\bibitem [{\citenamefont {Eschrig}\ \emph {et~al.}(2004)\citenamefont
{Eschrig}, \citenamefont {Richter},\ and\ \citenamefont
{Opahle}}]{Eschrig:2004wn}%
\BibitemOpen
\bibfield  {author} {\bibinfo {author} {\bibfnamefont {H.}~\bibnamefont
{Eschrig}}, \bibinfo {author} {\bibfnamefont {M.}~\bibnamefont {Richter}}, \
and\ \bibinfo {author} {\bibfnamefont {I.}~\bibnamefont {Opahle}},\ }in\
\href@noop {} {\emph {\bibinfo {booktitle} {Theoretical and Computational
Chemistry}}}\ (\bibinfo {year} {2004})\ p.\ \bibinfo {pages}
{723}\BibitemShut {NoStop}%
\bibitem [{\citenamefont {Enss}(2005)}]{Enss:2005ue}%
\BibitemOpen
\bibinfo {editor} {\bibfnamefont {C.}~\bibnamefont {Enss}},\ ed.,\ \href@noop
{} {\emph {\bibinfo {title} {{Cryogenic Particle Detection}}}},\ \bibinfo
{series} {Topics in Applied Physics}, Vol.~\bibinfo {volume} {99}\ (\bibinfo
{publisher} {Springer},\ \bibinfo {year} {2005})\BibitemShut {NoStop}%
\bibitem{Alotiby:2018}M. Alotiby \emph{et al.}, Phys. Med. Biol. \textbf{63} (2018) 06NT04 (6pp)
\bibitem{Koehler:2018}K.E. Koehler \emph{et al.}, J Low Temp Phys (2018) 193:1151–1159
\bibitem{Croce:2016}Croce, M.P., Rabin, M.W., Mocko, V. et al. J Low Temp Phys (2016) 184: 958.
\bibitem{Loidl:2018}M.Loidl, M.Rodrigues, R.Mariam, Applied Radiation and Isotopes
Vol. 134, (2018) 395-398 
\bibitem{Inoyatov:2015}A. Kh. Inoyatov \emph{et al.},Phys. Scr. \textbf{90}, 025402 (2015)
\bibitem{Inoyatov:2013} A.Kh. Inoyatov, L.L. Perevoshchikov, A. Kovalik, D.V. Filosofov, V.S. Zhdanov, A.V. Lubashevskiy, Z. Hons, Journal of Electron Spectroscopy and Related Phenomena, \textbf{187}, 61-64 (2013)
\bibitem{Inoyatov:2007}A. Inoyatov , D.V. Filosofov, V.M. Gorozhankin, A. Kovalik, L.L. Perevoshchikov, Ts. Vylov, Journal of Electron Spectroscopy and Related Phenomena, \textbf{160}, 54-57 (2007)
\bibitem{Inoyatov:2011}A. Inoyatov, A. Kovalik, D.V. Filosofov, L.L. Perevoshchikov,  V.S. Pronskih, Journal of Electron Spectroscopy and Related Phenomena, Vol. \textbf{184}, 52-56 (2011)
\bibitem{Inoyatov:2012}A. Kh. Inoyatov, L. L. Perevoshchikov, A. Kovalik, \emph{et al.} Eur. Phys. J. D \textbf{66}, 9, 234 (2012)
\bibitem{Tee:2019}B. P. E. Tee, A. E. Stuchbery, M. Vos, J.T.H. Dowie, B.Q. Lee, M. Alotiby, I. Greguric,  T. Kibedi, Phys. Rev. C \textbf{100}, 3, 034313 (2019)
\bibitem{Casey:1968}W.R. Casey and R. G. Albridge, Z. Physik \textbf{219}, 216--226 (1969)
\bibitem{Porter:1974}F.T. Porter, I. Ahmad, M.S. Freedman, J .Milsted, and A.M. Friedman, Phys. Rev. C \textbf{10}, 2, 803 (1974)
\bibitem{Vetter}P.A. Vetter \emph{et al.}, Phys. Lett. B \textbf{670}, 196-199 (2008)
\bibitem{Voytas:2002}P.A. Voytas, C. Ternovan, M. Galeazzi, D. McCammon, J.J. Kolata, P. Santi, D. Peterson, V. Guimaraaes, F.D. Becchetti, M.Y. Lee, T.W. ODonnell, D.A. Roberts, and S. Shaheen, Phys. Rev. Lett. \textbf{88}, 1, 012501 (2001)
\bibitem{Fontanelli:1996}F. Fontanelli, M. Galeazzi, F. Gatti, P. Meunier, A. Swift, S. Vitale , Nuclear Instruments and Methods in Physics Research A \textbf{370}, 273-275 (1996)
\bibitem{Meunier}P. Meunier, Nuclear Physics B (Proc. Suppl.) \textbf{66}, 207-209 (1998)
\bibitem{Hartmann:1985}F.X. Hartmann and R.A. Naumann, Phys. Rev. C \textbf{31}, 4, 1594 (1985)
\bibitem{Coron:2012}N. Coron, \emph{et al.}, Eur. Phys. J. A \textbf{48}:89 (2012)
\end{thebibliography}
\end{document}